\documentclass[11pt,a4paper]{article}   %,twocolumn
\usepackage{amsmath}
\usepackage{amssymb}
\usepackage{mathrsfs}

\usepackage{rotating}

\usepackage[round]{natbib}

\newcommand{\ie}{i.e.\ }

\newcommand{\real}{\textrm{Re}}

\newcommand{\wrt}{ ~ {\rm d}}

\def\ci{{\mathrm i}}
\newcommand{\ex}{{\rm e}}

\newcommand{\order}{\textrm{O}}

\usepackage[normalem]{ulem}
\usepackage{setspace}
\usepackage{xcolor}
\usepackage{marginnote}

\usepackage{xparse}

\NewDocumentCommand{\marcomm}{mo}
 {% #1 = text 
  \IfValueTF{#2}
  {\marginnote[\framebox{\parbox{40pt}{\setstretch{1.0}#1}}]{\framebox{\parbox{40pt}{\setstretch{1.0}#1}}}[#2] }
  {\marginnote[\framebox{\parbox{40pt}{\setstretch{1.0}#1}}]{\framebox{\parbox{40pt}{\setstretch{1.0}#1}}}[0pt] }
 }
 
 \NewDocumentCommand{\revmarcomm}{mo}
 {% #1 = text 
  \IfValueTF{#2}
  {\reversemarginpar\marginpar[\color{blue}\framebox{\parbox{33pt}{\setstretch{1.0}\scriptsize\sffamily#1}}]{\vspace{#2}\color{blue}\framebox{\parbox{33pt}{\setstretch{1.0}\scriptsize\sffamily#1}}} }
  {\reversemarginpar\marginpar[\color{blue}\framebox{\parbox{33pt}{\setstretch{1.0}\scriptsize\sffamily#1}}]{\vspace{-5pt}\color{blue}\framebox{\parbox{33pt}{\setstretch{1.0}\scriptsize\sffamily#1}}} }
 }
 
\begin{document}
\title{Hydrodynamic responses of a thin floating disk to regular waves}
\author{L.~J.~Yiew$^{1}$, 
L.~G.~Bennetts$^{1}$, 
M.~H..~Meylan$^{2}$, 
B.~J.~French$^3$, 
G.~A.~Thomas$^{3,4}$
\\
{\footnotesize
$^{1}$School of Mathematical Sciences, University of Adelaide, Adelaide, SA, Australia}
\\
{\footnotesize
$^{2}$School of Mathematical and Physical Science, University of Newcastle, NSW 2308, Australia}
\\
{\footnotesize
$^3$National Centre for Maritime Engineering and Hydrodynamics, Australian Maritime College, TAS 7250, Australia}
\\
{\footnotesize
$^4$Department of Mechanical Engineering, University College London, WC1E 6BT, U.K.}
}
\date{\today}
\maketitle

\begin{abstract}
Laboratory wave basin measurements of the surge, heave and pitch of a floating plastic disk caused by regular incident waves are presented.
The measurements are used to validate two theoretical   models: one based on slope-sliding theory and the other on combined potential-flow and thin-plate theories.  
\end{abstract}

% ############################################################################
\section{Introduction}

As ocean surface waves progress deeper into the partially sea ice covered ocean, they encounter discrete, relatively thin chunks of ice (floes) of increasing horizontal dimensions \citep{Squ&Mor80}. The range of dimensions depends on geographic location and season. However, the floes can be as small as a metre in the case of pancake ice, and up to hundreds of metres. The waves are attenuated by their interactions with the floes before they reach the quasi-continuous ice pack \citep{She&Ack91}, notwithstanding the large floes pushed into the outer fringes of the ice cover by random ice motions, which are subsequently broken up by the waves \citep{Squetal95}.

The waves impact the ice cover. They breakup the ice into smaller floes \citep{Pri&Pet11}, which are more prone to melting and easily stirred up by winds, currents and waves. For example, waves herd the floes into groups \citep{Wad83}. Further, waves cause floes to collide with one another \citep{Mar&Bec87},  which cause them to erode and produce rubble \citep{McK&Cro90}. The collisions can turn into rafting events, which can lead to floes bonding and hence thickness growth \citep{Daietal04}. Waves also introduce warm water and overwash the floes, which accelerates melt \citep{Wadetal79a,Mas&Sta10}. 

Theoretical/numerical models have been developed to predict wave impacts on the ice cover. \citet{She&Ack91} used a one-dimensional model to study collisions between floes and herding. They used the slope-sliding model of \citet{RumEA79} to calculate the horizontal motions of the floes induced by waves. The slope-sliding model is an extension of Morrison's equation, which includes a force due to the slope of the wave field. The model is derived on the assumption that floes do not modify the wave field, \ie the floe diameter is much less than the wavelength. It predicts the horizontal motion of a floe to be the sum of an oscillatory surge motion at the period of the incident wave, and a steady drift in the direction of the incident wave. \citet{SheZho01} derived analytical solutions to the slope-sliding model in certain cases. \citet{Mar99} independently derived a similar slope-sliding theory to \citet{RumEA79}. \citet{GroMey06} related the two theories and identified an error in the derivation of \citet{RumEA79}.

\citet{Koh&Mey08a} and \citet{Wiletal13a,Wiletal13b} modelled wave-induced breakup of a large group of ice floes. They applied breakup criteria that extended the earlier work of \citet{Lanetal01}. The kernel of both models is a model of a wave interacting with a solitary floe. The wave-floe interaction model uses linear potential-flow theory to model water motions and thin-plate theory to model the floe. The linear potential-flow/thin-plate model is commonly used to study wave-floe interactions \citep[see the review of][for example]{Squ07}. It assumes all motions are small perturbations from the equilibrium, \ie the floe oscillates but does not drift.

\citet{Koh&Mey08a} and \citet{Wiletal13a,Wiletal13b} used two-dimensional models (one horizontal dimension and one depth dimension). \citet{Mas&LeB89}, \citet{Meyetal97} and \citet{Benetal10} developed three-dimensional models of waves propagating through large groups of floes. They focussed on the attenuation of wave energy into the ice-covered ocean and did not model breakup or any other impact of the waves on the ice cover.  \citet{Mas&LeB89} and \citet{Meyetal97} modelled the floes using the thin-disk models of \citet{Isa82} and \citet{Mey&Squ96}, respectively, noting the former is a rigid model and the latter is an elastic model. \citet{Benetal10} used a disk model and also square-plate model, using the finite-element approach of \citet{Mey02}, but did not find the different shapes significantly altered the attenuation rates predicted.

\citet{BenWil15} recently used laboratory wave basin experiments to validate the model of \citet{Meyetal97}, and the two-dimensional model of \citet{Ben&Squ12a}, which was used by \citet{Wiletal13a,Wiletal13b}. They used arrays of 40 to 80 identical wooden disks to model the ice cover, and measured the proportion of wave energy it transmitted for regular incident waves over a range of wave frequencies and, in some cases, for two different amplitudes. The quotient of thickness, {$\tau$}, over diameter, $D$, for the disks was {$\tau/D\approx 3.3\times10^{-2}$ }. The quotient of the incident wavelength, $\lambda$, over the disk diameters was in the range $\lambda/D\approx 0.67$ to 6.28. The incident steepness, represented by the product $ka$, where $k=2\pi/\lambda$ is the wavenumber and $a$ is the incident amplitude, was in the range $ka\approx 0.04$ to 0.26. They showed the models predict the transmitted energy accurately for small incident amplitudes and low concentrations of the disks. They observed the models were inaccurate for the larger incident amplitudes when wave overwash of the disks was strong. (Overwash refers to the wave running over the top of the disks, due to their small freeboards.) Further, they provided evidence the models were inaccurate for high concentrations due to collisions between the disks, caused by out of phase surge motion of adjacent disks, and rafting, cause by out of phase heave and pitch motions. The potential-flow/thin-plate model does not include the highly nonlinear processes of overwash and collisions. 

Modelling collisions between disks requires an accurate model of the surge motion of a solitary disk. Heave and pitch motions must also be modelled to predict rafting. However, the potential-flow/thin-plate and slope-sliding model predictions of these oscillatory motions have not yet been thoroughly validated. In particular, the solitary disk experiences overwash for moderate incident amplitudes, which is not contained in either model. 

\citet{BenWil15} presented measured surge, heave and pitch motions of a solitary wooden disk for a subset of the incident frequencies and amplitudes used for their multiple-disk tests. They compared the measurements to the predictions of the potential-flow/thin-plate model, and found the model is, in general, accurate. However, they also showed the model was least accurate for a test in which strong overwash occurred. In particular, the model overpredicted the translational motions, surge and heave, and underpredicted the rotational motion, pitch.

Previously, \citet{Monetal13a,Monetal13b} presented measurements of the oscillatory motions of a thin plastic disk in response to regular incident waves, as functions of the incident frequency. They used three thin disks, with $\tau/D=2.1\times 10^{-3}$ to $6.9\times 10^{-3}$, and incident waves with lengths ranging from $\lambda/D\approx 0.63$ to 3.14, and two small steepnesses $ka\approx 0.03$ and 0.06. They compared the measurements to predictions of the potential-flow/thin-plate model. However, they focussed on the flexural motion of the disk. They used a vertical rod through the centre of the disk to suppress surge, and a barrier around the edge of the disk to prevent overwash.

\citet{Meyetal15a} presented measurements of the surge, heave and pitch motions of a thin plastic disk, as functions of $\lambda /D$. They used a disk with thickness over diameter quotient $\tau/D\approx 3.8\times10^{-2}$,
and incident waves with lengths ranging from $\lambda/D\approx 0.9$ to 12.3 and steepness ranging from $ka\approx 0.01$ to 0.3. They compared the surge measurements to predictions of the slope-sliding model. They showed the model predictions are accurate for incident wavelengths approximately greater than two floe diameters, for suitably chosen model parameters. However, they also used a barrier around the edge of the disk to prevent overwash.

\citet{McG&Bai14} presented measurements of the heave and a composite surge and drift of model floes made of paraffin wax. They tested a variety of shapes, including square, rectangular and triangular shapes, but not disks. They modelled thick multiyear floes, and hence used relatively large thickness over characteristic length, $D_{c}$, quotients, typically $\tau/D_{c}=\order(10^{-1})$. They used regular incident waves with lengths in the range $\lambda/D_{c}=0.1$ to 0.75, and steepness in the range $ka\approx 0.03$ to 0.28. They studied heave and composite surge-drift as functions of $\lambda/D_{c}$ and $2a/\lambda$, but did not compare these results to model predictions. They noted the occurrence of overwash for large incident amplitudes and high steepnesses, and suggested this may be the source of the reduced heave responses they found in this regime, which mirrors the finding of \citet{BenWil15}.

\citet{McG&Bai14} also presented measurements of the drift of their model floes. They compared the measurements to the predictions of Stokes drift theory. They found the theory slightly underestimates the measurements. This finding is consistent with that of \citet{Huaetal11}. Laboratory experimental studies of the drift of floes have also been conducted by, for example, \citet{Har87} in a two-dimensional setting. 

In this study, surge, heave and pitch motions are extracted from an extended dataset to that used by \citet{Meyetal15a}. The extended dataset includes motions of a disk without an edge barrier, thus allowing inferences about the effect of overwash on the motions to be gained.  The motions are presented as functions of $\lambda/D$ for three incident wave amplitudes. The motions for the disk without the barrier are shown to be consistently smaller than those with the barrier, especially for heave in the short-wavelength regime. Surge, which is the focus of this study due to its relevance to collisions, is also presented as a function of steepness, $ka$, for two incident wavelengths. The surge response is shown to decrease as steepness increases for the disk without a barrier and the shorter wavelength but is approximately constant otherwise. %\lu{rephrase maybe? isn't this more of a summary than an intro?}

Further, the surge motions predicted by slope-sliding and potential-flow/thin-plate models are compared analytically and numerically in the long-wavelength regime. They are shown to have the same form. The surge, heave and pitch motion predictions of the potential-flow/thin-plate model are compared to the experimental measurements, as a function of $\lambda/D$. In the short-wavelength regime, the model is shown to agree better with the data for the disk without a barrier.

% ############################################################################
\section{Experiments}

\subsection{Method} \label{sec:Exp_method}

\begin{figure}
  \centerline{\includegraphics[width=0.6\textwidth]{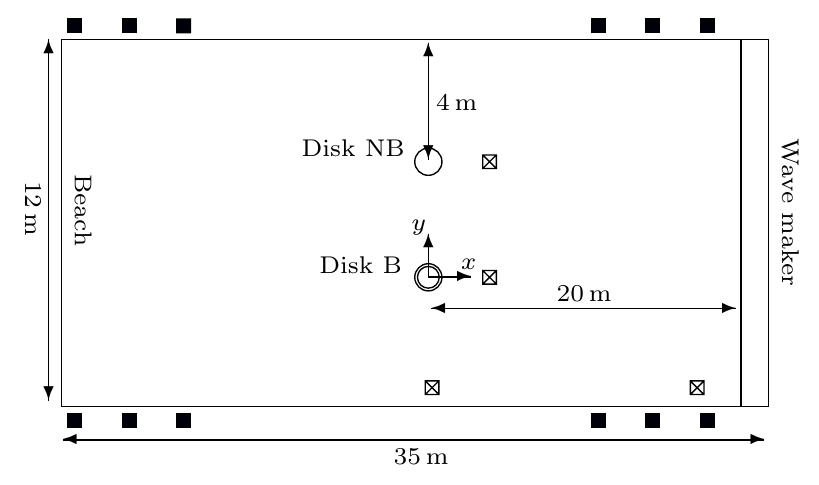}}
  \caption{Schematic plan view of the \textsc{mtb}. Wave probes ($\boxtimes$) record the wave profiles at set locations, and motion tracking cameras ($\blacksquare$) record the motions of the disks.}
  \label{fig:Exp_mtb}
\end{figure}

Laboratory experiments were performed at the Australian Maritime College, Launceston, Australia, using the Model Test Basin (\textsc{mtb}) facility. 
During the experiments, the wave induced motions of solitary floating disks  were recorded.
Figure \ref{fig:Exp_mtb} shows the plan view of the \textsc{mtb} and experimental set-up.

% ========================================================================
%\subsection{Experimental Setup}

The \textsc{mtb} is 35\,m long and 12\,m wide. 
It was filled with fresh water of density $\rho = $1000\,kg/m$^3$ to a depth of $h=0.83$\,m.
A piston-type wave maker bounds the \textsc{mtb} at its right-hand end.
A sloping beach bounds the \textsc{mtb} at its left-hand end. 
%\marcomm{Change all depths to $H$. \\ Luke: I'll leave depth as $h$, see comment below for thickness.}

Locations in the \textsc{mtb} are defined by the Cartesian coordinate system $(x,y,z)$.
The coordinate $(x,y)$ defines locations in the horizontal plane, parallel to the equilibrium water surface.
The coordinate $x$ points from the beach to the wave maker.
The coordinate $z$ defines the vertical location. It points upwards and its origin is set to coincide with the equilibrium water surface.

Tests were conducted for a range of regular incident wave conditions. 
Target wave amplitudes from $a= 2.5$\,mm to 50\,mm,  
and frequencies from $f=0.5$\,Hz to 2.0\,Hz, were tested. 
The corresponding wavelengths were approximately $\lambda=$ 0.4\,m to 5\,m.
The wavelengths are calculated as $\lambda=2\pi/k$, where the wave number $k$ is the positive real root of the dispersion relation 
\begin{equation} \label{eq:dispersion}
k \tanh{kh} = \kappa
\qquad
\textrm{where}
\qquad
\kappa = \frac{\omega^2}{g}
\end{equation}
is a frequency parameter,  $\omega = 2 \pi f$ is angular frequency, and $g\approx 9.81$\,m\,s$^{-2}$ is gravitational acceleration.

Four wave probes were installed around the wave basin to record the generated wave properties.
The measured incident frequencies closely matched the target values.
The measured incident amplitudes were generally slightly smaller than the target amplitudes.
The results presented in \S\,\ref{sec:ExpRes} to \S\,\ref{sec:Res} use the target frequencies and the measured amplitudes.

Tests were conducted for two matrices of incident wave amplitudes and frequencies.
The first matrix contained more frequency entries than amplitude entries.
The second matrix contained more amplitude entries.
Table~\ref{tab:Exp_conditions} summarises the tests conducted.
Tests were not conducted for the largest amplitude and highest frequency combinations, 
to avoid wave breaking.

\begin{table}
\includegraphics[width=1.0\textwidth]{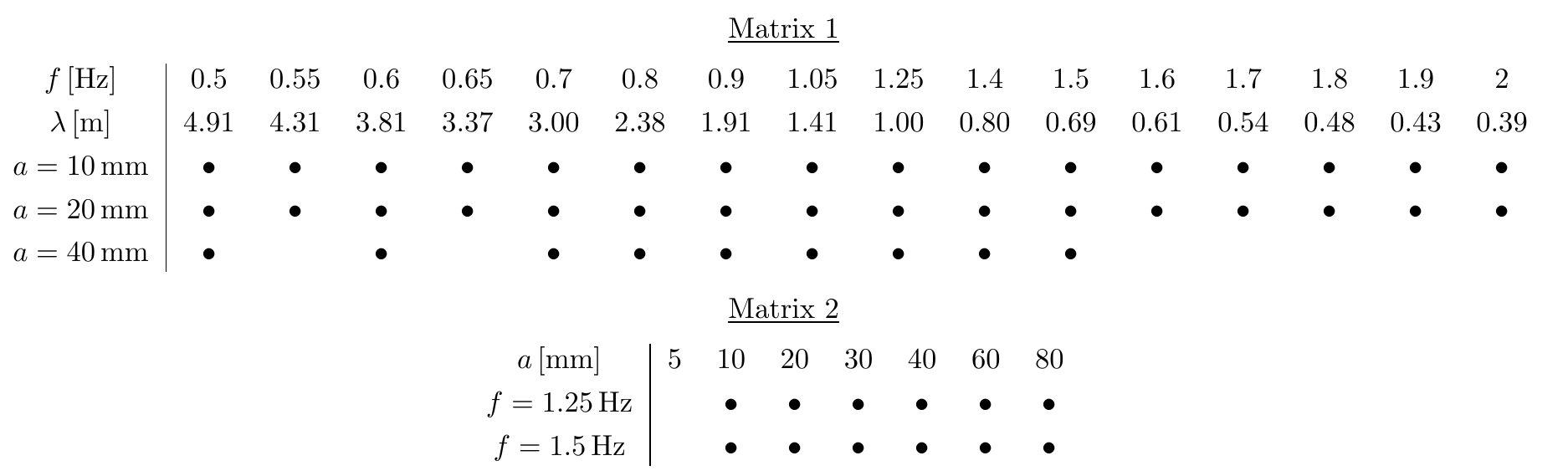}
\caption{Summary of test matrices.} 
\label{tab:Exp_conditions}
\end{table}

Two thin plastic Nycel disks were installed in the \textsc{mtb}. 
Nycel is an expanded rigid foam \textsc{pvc}.
The disks had radii $R=200$\,mm, thickness {${\tau=}15$\,mm}, density $\rho_{d}\approx 636$kg\,m$^{-3}$,
and hence equilibrium draft $d\approx 9.55$\,mm and mass $m=1.2$\,kg.
%\marcomm{Change all thicknesses to $D$. \\ Luke: I'll change all thicknesses to $\tau$ since $D$ is already used for diameter}[-20pt]

\begin{figure}
  \centerline{\includegraphics[width=\textwidth]{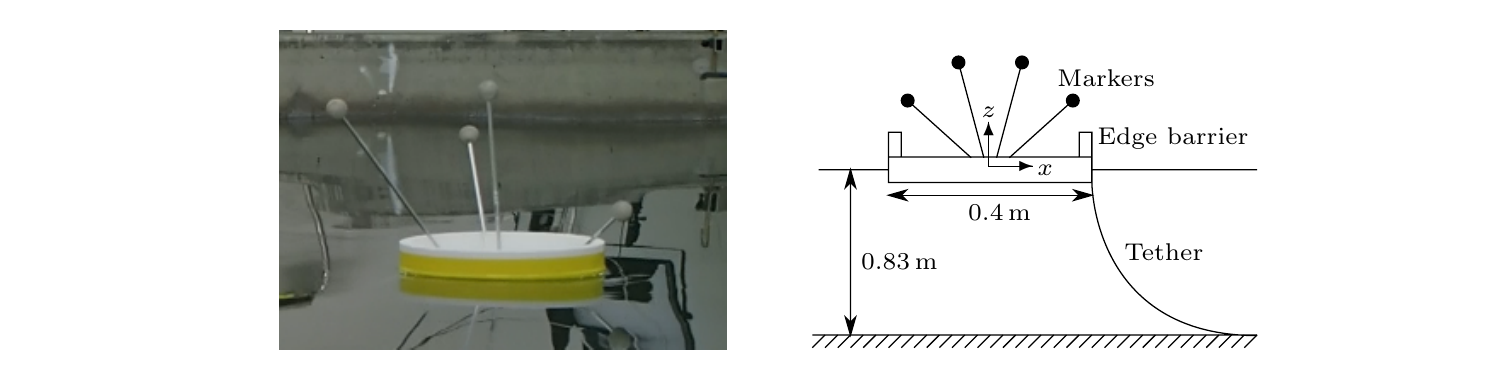}}
  \caption{Left-hand panel: photo of Disk~B. Edge barrier, and rods and markers are visible. 
  Right-hand panel: cross-sectional schematic of Disk~B, which shows the mooring system.}
  \label{fig:Exp_floe2}
\end{figure}

An edge barrier was attached to one disk, which is referred to as Disk B. 
The disk with no barrier is referred to as Disk NB.
The left-hand panel of figure~\ref{fig:Exp_floe2} shows a photo of Disk B. 
The edge barrier is a 50\,mm high and 25\,mm thick styrofoam ring.
The barrier is used to prevent waves overwashing the surface of the disk, due to its small freeboard,
and hence investigate whether the overwash affects disk motions.

The locations of the centres of mass of the the disks at time $t$ are denoted $(x,y,z)=(X(t),Y(t),Z(t))$.
The two disks were initially positioned approximately halfway down the basin.
They were placed 4\,m apart to minimise their scattered waves interfering with one another.
Similarly, they were each placed 4\,m from the \textsc{mtb} side walls, to minimise the effect of their scattered waves 
being reflected back to them.
The origin of the Cartesian coordinate system in the horizontal frame is set to coincide with the geometric centre of
Disk~B in the horizontal plane, and, thus, $(X(0),Y(0))=(0,0)$ for Disk~B and $(X(0),Y(0))=(0,4)$ for Disk~NB. 
The disks were anchored to the floor via loose elastic tethers to prevent them from drifting too far down the tank and to assist in resetting the initial positions after each test.

Four light-weight tracking balls were attached to each disk via aluminium rods.
Figure \ref{fig:Exp_floe2} shows the markers and rods on Disk~B.
The position of each marker was captured by the Qualisys non-contact motion tracking system. 
The system consists of eight pairs of infrared cameras and receivers installed along the perimeter of the wave basin. 
Qualisys records the locations of each marker in real-time at 200 frames per second. 
It uses this information to calculate the translational and rotational motions of the disks.
The markers were placed at different heights to minimise the chance of them overlapping in the camera image.
Data was collected from $t=0$, when the wave maker was activated, up to $t=60$\,s, when the wave maker was turned off.

\begin{figure}
  \centerline{\includegraphics[width=\textwidth]{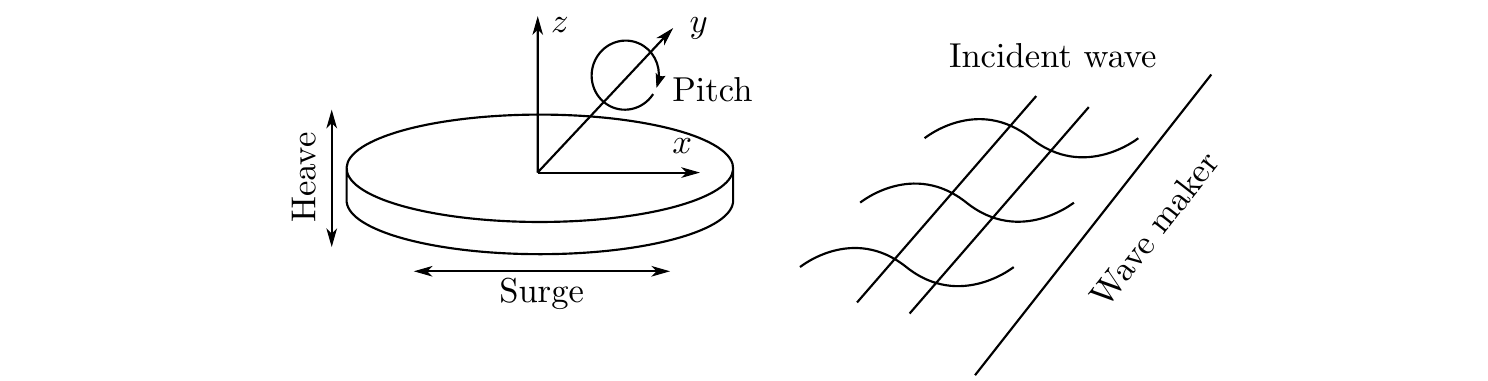}}
  \caption{Schematic of Disk~NB tests, including coordinate system and oscillatory motions.}
  \label{fig:Exp_coord}
\end{figure}

%\revmarcomm{Rewrite para: use coordinates \& note drift}[8pt]
Motions in the six rigid degrees of freedom of the disks were recorded. 
The symmetric motions, with respect to the $x$-axis, were dominant due to the plane incident waves.
%The symmetric motions consist of two translational motions
%and one rotational motion.
The translational motion in the $x$-direction is a combination of drift, restoration due to the mooring system, and surge, which is an  oscillatory motion at the frequency of the incident waves.
The translational motion in the $z$-direction is oscillatory heave.
The rotational motion in the $xz$-plane is oscillatory pitch.
%{Surge and heave are defined as the oscillatory motions in the $x$ and $z$ axes, respectively, and pitch coincides with rotations in the $xz$ plane. 
Figure~\ref{fig:Exp_coord} illustrates the oscillatory motions, which are analysed in this investigation.

% ========================================================================
\subsection{Data processing}

\begin{figure}
  \centerline{\includegraphics[width=\textwidth]{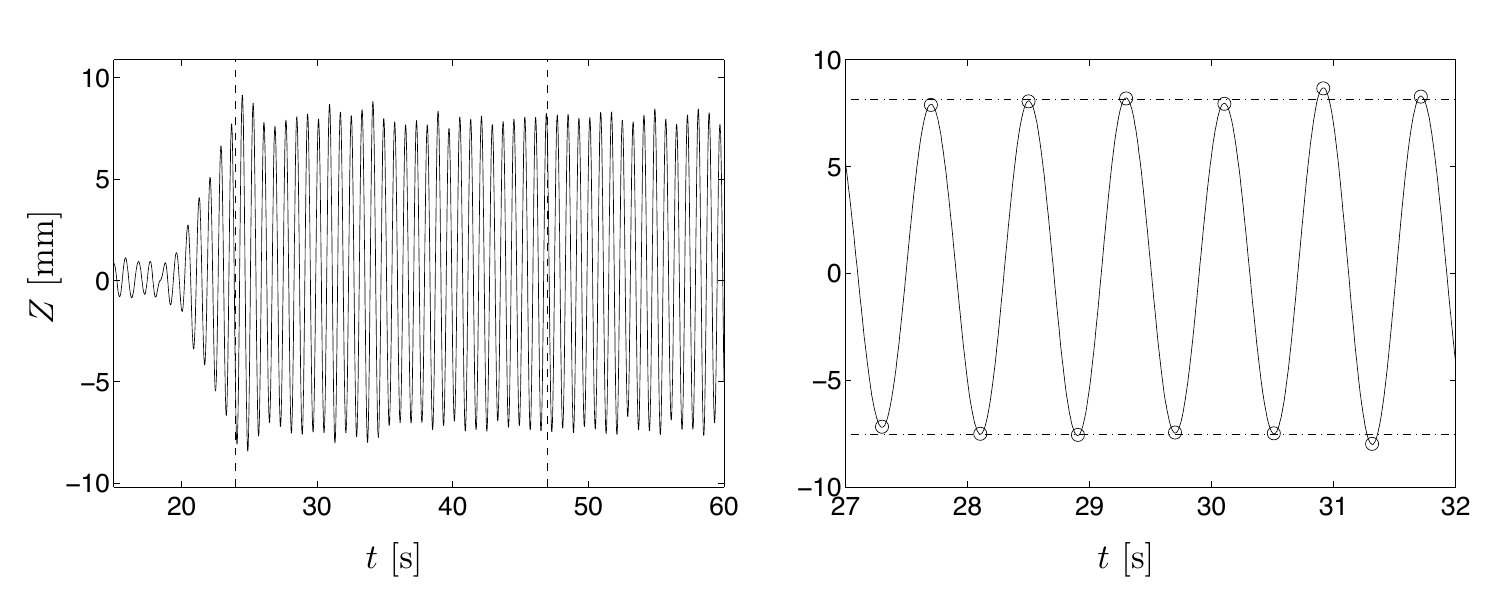}}
  \caption{Left-hand panel: example motion in $z$-direction. Steady-state interval is denoted by dashed vertical lines.  
  Right-hand panel: close-up of smoothed signal (solid curve). Local maxima and minima are denoted by circles. %($\circ$). 
  Mean maxima and minima are denoted by dot-dashed lines.}
  \label{fig:Exp_heave}
\end{figure}

{Figure \ref{fig:Exp_heave} shows an example time series provided by Qualisys
for the translational motion of Disk B in the $z$-direction. 
The test shown used an incident wave with frequency $f=1.25$\,Hz and measured amplitude $a = 8.5$\,mm.
%\lu{Which disk?}

Motions are considered in the steady-state interval only, which begins after the initial transient phase of the motions pass, and ends 
when waves reflected by the beach begin to interfere with the motions. 
The left-hand panel of figure \ref{fig:Exp_heave} denotes the intervals with vertical dotted lines.
The interval is calculated using the phase velocity of the wave, $c = f \lambda$. 

The \textsc{matlab} \texttt{smooth} function with the \texttt{lowess} method is used to eliminate noise from the raw signal and
clearly identify the local maxima and minima.
The \texttt{smooth} function applies a local regression method using a weighted linear least squares algorithm. 
The degree of smoothing is controlled by the smoothing parameter --- a factor of 0.01 is specified here. 
The right-hand panel of figure~\ref{fig:Exp_heave} shows the smoothed signal for a 5\,s interval at the beginning of the steady-state interval.

The maxima and minima are identified, as shown by the circles in the right-hand-panel of figure \ref{fig:Exp_heave}. 
The heave amplitude, $a_{H}$, is calculated as half the difference between the average peak and trough values.
%Figure \ref{fig:Exp_heave}\,(b) demonstrates this. 
The same procedure is used to calculate the pitch amplitude, $a_{P}$, from the time series of the translational motion.

\begin{figure}
  \centerline{\includegraphics[width=\textwidth]{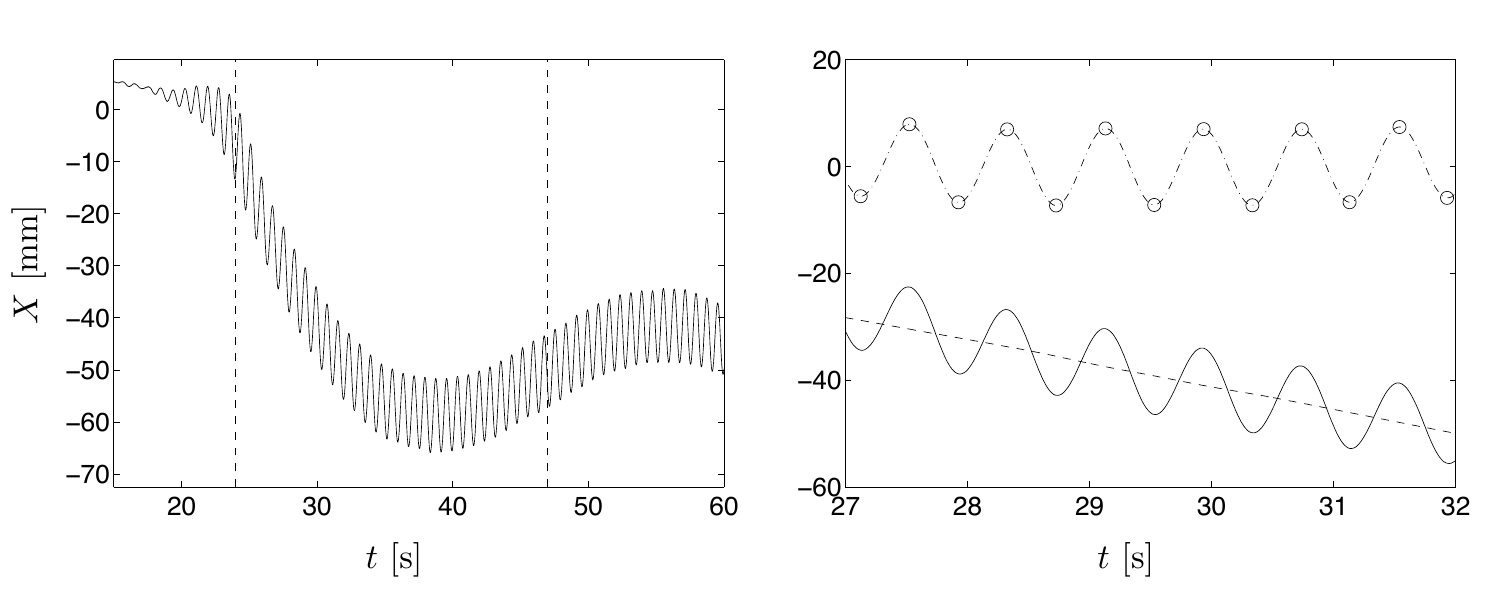}}
  \caption{Left-hand panel: as in left-hand panel of figure~\ref{fig:Exp_heave} but for motion $x$-direction. 
  Right-hand panel:  close-up of smoothed motion (solid curve), and decomposed into surge (dot-dashed) and drift (dashed).  Local maxima and minima of surge motion are denoted by circles.}
  \label{fig:Exp_surge}
\end{figure}

The left-hand panel of figure~\ref{fig:Exp_surge} shows the translational motion in the $x$-direction, 
for the same test considered in figure~\ref{fig:Exp_heave}.
Surge and drift are visible in the motion, as is the long-period restoration due to the mooring system.

The \texttt{smooth} function is used with a smoothing factor of 0.9 %\lu{what?} 
to extract the drift from the Qualysis data.
The surge motion is the difference between the full data and the drift.
The right-hand panel of figure \ref{fig:Exp_surge} illustrates the decomposition for the small time interval.
The surge amplitude, $a_{S}$, is, subsequently, calculated using the same procedure as for heave and pitch.}

% ========================================================================
\subsection{Response amplitude operators}\label{sec:ExpRes}

\begin{figure}
  \centerline{\includegraphics[width=\textwidth]{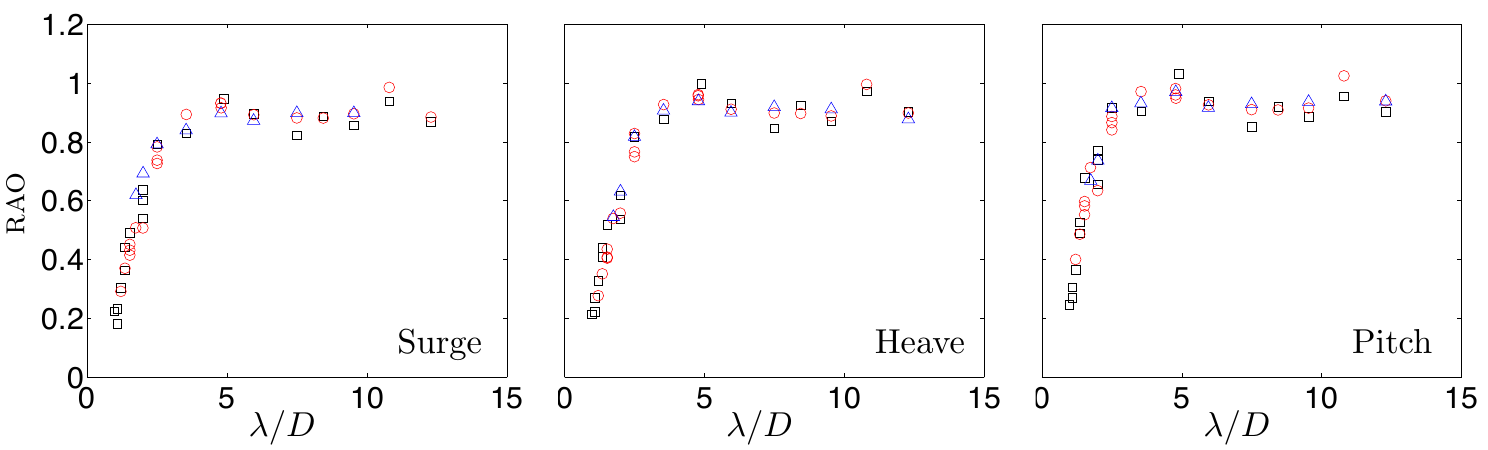}}
  \caption{\textsc{Rao}s for Disk~NB, as functions of nondimensional incident wavelength.
   Data are grouped according to incident wave amplitude: $a=$10\,mm ({$\square$}), 20\,mm (\textcolor{red}{$\circ$}) and 40\,mm (\textcolor{blue}{$\bigtriangleup$}).}
  \label{fig:Exp_Res_Lin1}
\end{figure}

{Figure \ref{fig:Exp_Res_Lin1} shows response amplitude operators (\textsc{rao}s) of Disk~NB for surge, heave and pitch, as functions of the incident wavelength nondimensionalised with respect to the disk diameter, $D=2R$. 
The results are for the first test matrix.
The surge \textsc{rao} is $a_{S}/\{a\coth{kh}\}$, 
where $a\coth{kh}$ is the horizontal extent of the trajectory of a fluid particle at the free surface. 
The heave \textsc{rao} is $a_{H}/a$.
The pitch \textsc{rao} is $a_{P}/ka$, where $ka$ is the incident wave steepness.
%\lu{Perhaps put the two figures together.}
%Results for Disk~B ... blah blah blah ...
In the sub-panels, data are grouped according to incident wave amplitude (using different symbols and colours).
Error bars are omitted here for clarity. They are presented in figures \ref{fig:Res_PFvExp_surge} to \ref{fig:Res_PFvExp_pitch}.

%Fig.~\ref{fig:Exp_Res_Lin1} shows \textsc{rao}s for Disk~B.
%Fig.~\ref{fig:Exp_Res_Lin2} shows \textsc{rao}s for Disk~NB.
The \textsc{rao}s are independent of the incident wave amplitude.
The mean range of the corresponding \textsc{rao}s for the different amplitudes is approximately 0.067.
Moreover, no behavioural trend with varying amplitude is evident.
%After normalising the surge, heave and pitch amplitudes with respect to $a$, the data collapses into a single curve, as seen in Figs.\ \ref{fig:Exp_Res_Lin1} and \ref{fig:Exp_Res_Lin2}. 
The results suggest a linear relationship exists between the amplitudes of oscillatory motion and the incident wave amplitude.
%Figure ... demonstrates the linearity...
 
\begin{figure}
  \centerline{\includegraphics[width=\textwidth]{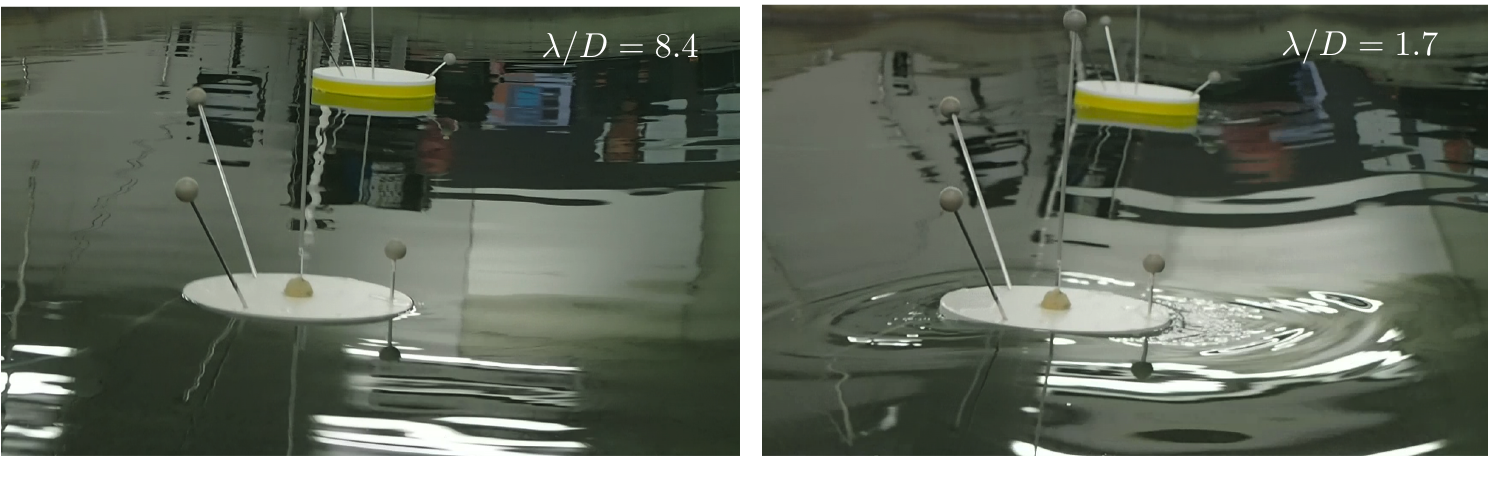}}
  \caption{Example of changes to the wave field in the long- and short-wavelength regime (left-hand and right-hand panels, respectively). Scattering is evident in the short-wavelength regime.}
  \label{fig:Exp_overwash}
\end{figure} 
 
The \textsc{rao}s are approximately unit value for $\lambda / D$ greater than 3.
The disk does not affect the incident waves in this long-wavelength regime. 
The left-hand photo in figure \ref{fig:Exp_overwash} shows an example of this behaviour.
%\lu{Figures must be in the order to which they are referred in the text!
%Also, swap the photos left-to-right.}
The disk simply follows the path of a fluid particle at the free surface. 
%\citep{MeyEA14,Mar99,MasLeB89}.

The \textsc{rao}s decrease as the incident wavelength decreases.
%In the short wavelength regime, the motions begin to decrease with wavelength. 
The disks scatter the incident waves in the short-wavelength regime. 
The right-hand photo of figure \ref{fig:Exp_overwash} shows an example of this. 
Scattering results in less energy being transferred into the oscillatory motions,
and thus reduces the \textsc{rao}s.

\begin{figure}
  \centerline{\includegraphics[width=\textwidth]{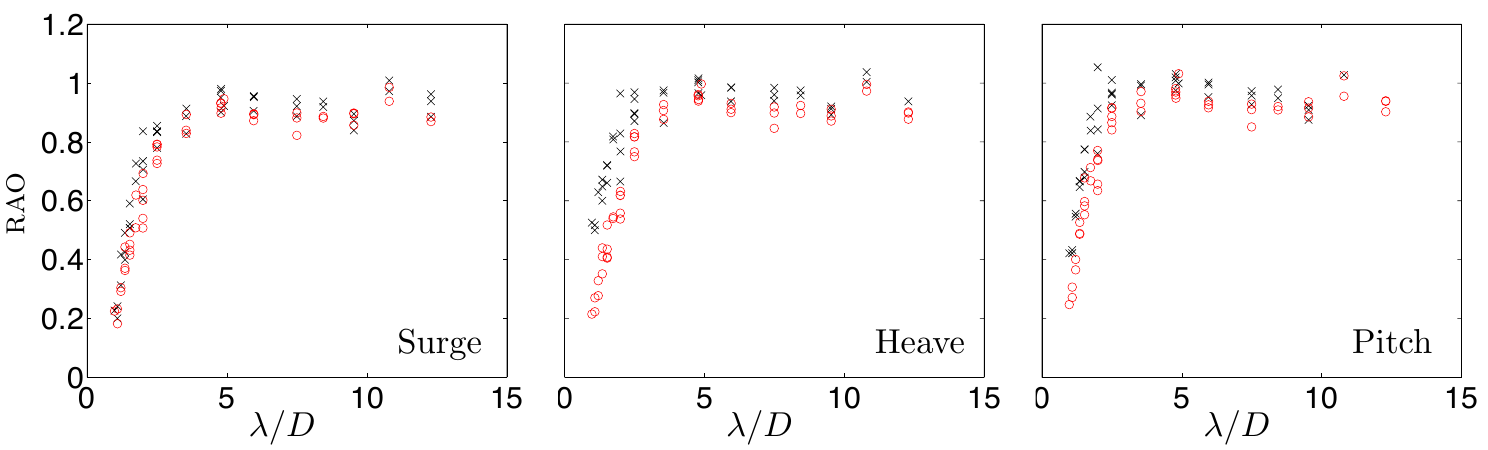}}
\caption{\textsc{Rao}s for Disk~B ($\times$) and Disk~NB (\textcolor{red}{$\circ$}).}
\label{fig:Exp_Res_Ice1Ice2}
\end{figure}

Figure \ref{fig:Exp_Res_Ice1Ice2} compares the \textsc{rao}s of the two disks for the tests in the first matrix. 
Tests with different incident amplitudes are not distinguished from one another.
Disk~NB generally has slightly smaller \textsc{rao}s than Disk~B. 
The difference is greatest for heave at 14.8\% and least for surge at 7.8\%.
The percentages quoted represent the mean of the 
difference in mean \textsc{rao} for the two disks,
for an incident wavelength and \textsc{rao} combination.
Differences are most pronounced in the short-wavelength regime, $\lambda/D < 3$.
The largest difference is 0.35, which occurs for heave at $\lambda/D = 0.9$.

\begin{figure}
  \centerline{\includegraphics[width=\textwidth]{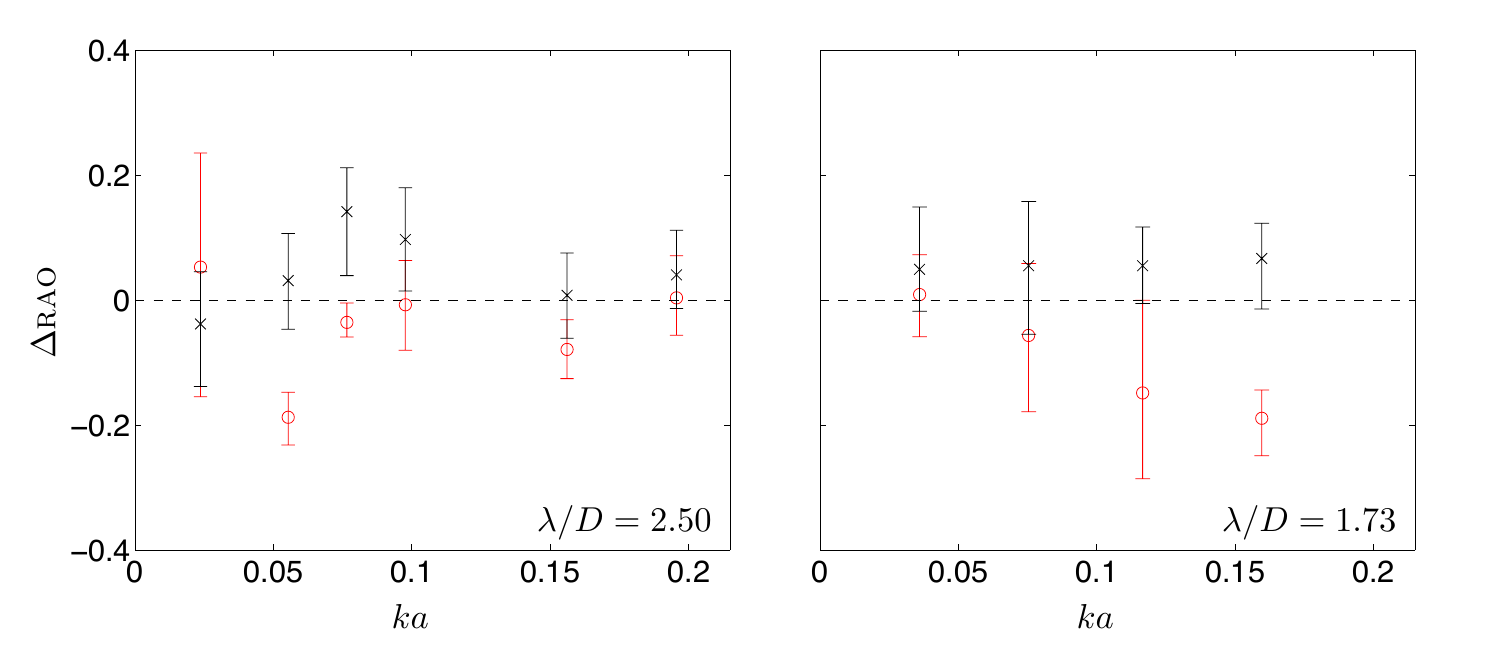}}
\caption{Deviation of surge \textsc{rao} from the mean, as a function of wave steepness, for incident frequency $f = 1.25$ (left-hand panel) and 1.5 (right). Results for Disk~B ($\times$) and Disk~NB (\textcolor{red}{$\circ$}). Error bars indicate deviation of the maxima/minima.}
\label{fig:Exp_RAO_Lin}
\end{figure}

Figure~\ref{fig:Exp_RAO_Lin} shows the surge \textsc{rao}, as a function of the incident wave steepness, $ka$, 
calculated from tests in the second matrix.
%the surge \textsc{rao}s as functions of the incident wave steepness, $ka$, calculated from .
Surge motion is presented here, as it is the focus of this study.
Cognate behaviours occur for heave and pitch motions (not shown). 
%\marcomm{Lucas: have you checked this? \\ Luke: Yes. Identical.} 

The results are grouped according to the incident wavelength used (different panels)
and which disk the results refer to (different symbols and colours).
The \textsc{rao}s are presented in terms of their relative deviation from the mean, defined by
\begin{equation}
\Delta\textrm{\textsc{rao}} = \frac{\textrm{Surge~\textsc{rao}}-\langle \textrm{Surge~\textsc{rao}} \rangle}{\langle \textrm{Surge~\textsc{rao}} \rangle},
\end{equation}
where angled brackets $\langle\cdot\rangle$ denote the mean with respect to both disks and different steepnesses.
The error bars denote \textsc{rao} values for the  maximum and minimum surge amplitudes in the tests.
%\marcomm{Lucas: is (2.2) correct? \ie is the deviation relative? Yes.}[0pt]

For the longest incident wavelength tested, $\lambda/D=2.50$, the \textsc{rao} values generally {deviate} by no more than  10\% from the mean.
The deviations here do not display a consistent trend and are, therefore, attributed to measurement errors.
For the shortest wavelength tested, $\lambda/D=1.73$, the \textsc{rao}s for Disk~B are insensitive to steepness.
In comparison, the \textsc{rao}s for Disk~NB decrease as steepness increases.
This is attributed to the wave overwashing the Disk~NB in this regime, 
and hence suppressing surge.
This finding is consistent with that of \citet{BenWil15}.

% ############################################################################
% ############################################################################
\section{Theoretical Models}

\subsection{Slope-sliding model} \label{sec:SS}

%The slope sliding model presented by \citet{Gro&Mey06} and \citet{Meyetal15a} is summarised here for convenience.
The slope-sliding model assumes the floating body, here the disk, moves along the wave profile due to gravity.
Its movement is resisted by drag between the body and water.
The equation of motion %of the floe 
in the $x$-direction is thus
\begin{equation} \label{eqn:SS}
m(1+c_m) 
\frac{d V}{ d t}= 
- m g
 \left[
 \frac {\partial \eta} {\partial x} 
 \right]_{x=X}
+ 
\rho 
c_d 
l
|\widehat{V}| 
\widehat{V}.
\end{equation}
This nonlinear ordinary differential equation %(\ref{eqn:SS}) 
is solved for the horizontal velocity of the disk, $V(t)=dX(t)/dt$.

The term on the left-hand side of equation (\ref{eqn:SS}) is the inertial force of the disk.
The coefficient $c_m$ is the added mass of the disk, \ie its increased resistance to motion due to contact with water.
The first term on the right-hand side is the sliding force due to gravity.
The quantity $\eta(x,t)$ is the wave profile.
The second term is the drag force. 
%The quantity $V_w(x,t)$ is the velocity of water particles. 
The quantity $\widehat{V}=V_{w}(t)-V(t)$, where $V_w(t)$ is the velocity of water particle on the free surface at $x=X$. 
%is the floe velocity relative to the velocity of the water particle at $x$ and time $t$.
The coefficient $c_d$ is the drag coefficient and $l=\pi R^{2}$ is the wetted surface area of the disk.

The model further assumes the disk diameter is small in relation to the incident wavelength.
As discussed in \S~\ref{sec:ExpRes}, in this regime the incident wave profile is not modified by the floe, \ie
\begin{equation}\label{eqn:wp}
\eta(x,t)
 = 
a
\sin( k x - \omega t ).
\end{equation}
The phase of the incident wave does not affect the \textsc{rao}s extracted from the model, 
and it is therefore normalised to the origin.

Using the wave profile (\ref{eqn:wp}) and the notation $Q=kX-\omega t$, 
equation (\ref{eqn:SS}) reduces to the autonomous dynamical system
\begin{subequations}\label{eqn:SS2} 
\begin{equation} \label{eqn:SS2a} 
\frac{d V}{ d t} 
= 
\frac{
%\rho C_d W \left | {\pi f H \sin\theta - V} \right | 
%\left ( \pi f H \sin\theta - V \right )
- mg ka \cos Q
+
\rho c_d l |\widehat{\mathcal{V}}| \widehat{\mathcal{V}}
}{m(1+c_m)}
, 
\end{equation}
where $\widehat{\mathcal{V}}(V,Q)=\frac12 \omega a \sin Q - V$, and
\begin{equation}
%\textrm{and}
%\qquad
\frac{d Q}{dt}= k V - \omega.
\end{equation}
\end{subequations}
Figure~\ref{fig:SS_phase} shows the phase plane for an example problem, 
with zero drag and added mass, $c_{d}=c_{m}=0$,
and an incident wave of amplitude $a=50$\,mm and length $\lambda=6$\,m.

System~(\ref{eqn:SS2}) always has a limit cycle.
Figure~\ref{fig:SS_phase} illustrates the limit cycle with the solid (red) curve that intersects the origin.
The limit cycle represents steady surge (the amplitude of the cycle) and drift (the difference in period of the cycle and the incident wave period) of the disk. 
This is the relevant solution for the present investigation.

The example problem also possesses two fixed points at $(Q,V)=(\pi/2,\omega/k)$ and $(3\pi/2,\omega/k)$.
These are the so-called surfing solutions.
They can only be reached for a large initial velocity, $V(0)$.
The right-hand fixed point is a centre, \ie the trajectories surrounding it are closed.
The left-hand fixed point is a saddle node, which has lower and upper homoclinic connections.
For all initial conditions below its lower homoclinic connection in the phase plane, 
the orbits tend to the limit cycle.

For small non-zero values of the drag coefficient, the fixed points are shifted and 
the right-hand fixed point becomes a stable spiral.
However, the system retains the structure that all orbits below the lower homoclinic connection tend to the limit cycle.
For large values of the drag coefficient, the fixed points disappear in a saddle-node bifurcation.
All orbits then tend to the limit cycle.

In practice, 
the limit cycle is obtained by solving system~(\ref{eqn:SS2}) as the second-order ordinary differential equation
\begin{equation}\label{eqn:SS2b}
m(1+c_m) 
\frac{d^{2} X}{ d t^{2}} 
+ 
m g
ka \cos(kX-\omega t) 
- 
\rho 
c_d 
w
|\widehat{V}| 
\widehat{V}
=
0,
\end{equation}
with zero initial displacement and velocity, \ie $X=dX/dt=0$ for $t=0$.
The solution to (\ref{eqn:SS2b}) is obtained numerically using the \textsc{matlab} package \texttt{ode45} solver, which is based on a fourth and fifth-order Runge-Kutta method.
The steady solution, \ie the limit cycle, is decomposed into 
surge and drift components, as for the experimental data.
%Surge amplitudes are extracted from the steady-state solution.
%\marcomm{Lucas: correct statement re RK}[-40pt]

\begin{figure}
  \centerline{\includegraphics[width=\textwidth]{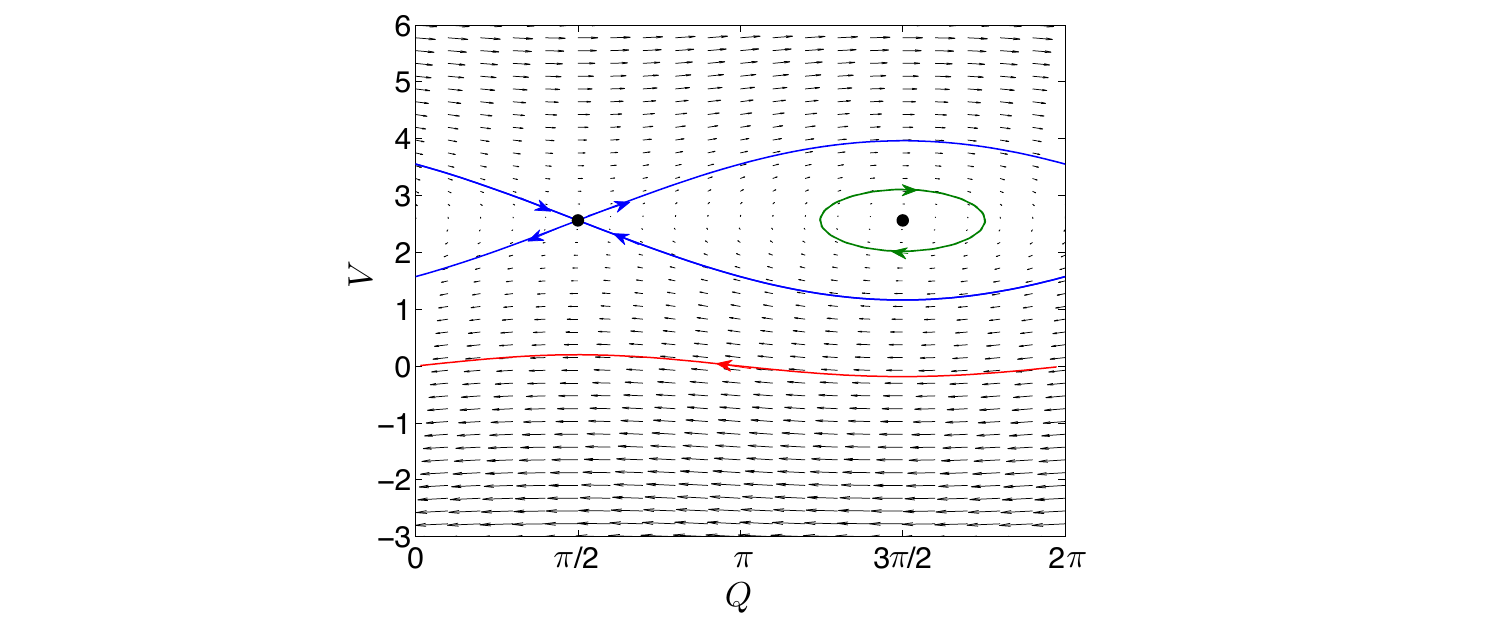}}
  \caption{Phase diagram for horizontal disk velocity, $V$,  and phase, $Q$, for example problem with 
  $\omega = 2.68$,
  $c_m=c_d = 0$, $a = 0.05$\,m and $\lambda = 6$\,m.
%  \lu{What is the value of $\omega$?}
  Vectors show the evolution of $V$ and $Q$. 
  Here two equilibrium points exist at $(1.57,2.56)$ and $(4.71,2.56)$. The lower red curve represents the limit cycle.}
  \label{fig:SS_phase}
\end{figure}

The values of the added mass and drag coefficients are obtained via 
comparison with experimental data.
\citet{MeyEA14} compared the surge \textsc{rao} obtained from the slope-sliding model 
to the experimental surge data for Disk~B.
They found the parameters $c_m=0.1$ and $c_d=0$ provided a good fit to the data for $\lambda/D>2$,  approximately.
For shorter incident wavelengths, the slope-sliding model does not capture the surge for any choice of added mass and drag coefficients.

%%%%%%%%%%%%%%%%%%%%%%%%%%
%%% POTENTIAL FLOW MODEL %%%
%%%%%%%%%%%%%%%%%%%%%%%%%%

\subsection{Linear potential-flow/thin-plate model}\label{sec:PF}

The linear potential-flow/thin-plate model assumes the disk oscillates about its equilibrium position.
The oscillations are time harmonic at the frequency of the incident wave,
and are symmetric with respect to the $x$-axis, \ie in-line with the incident wave.
The location of the centre of mass of the floe in the $x$-direction is denoted
\begin{equation}
X(t)=\real\{A_{S}\ex^{-\ci \omega t}\},
\end{equation}
where $A_{S}$ is the complex-valued surge amplitude, 
such that its magnitude is the surge amplitude, $a_{S}$, and its argument is the phase of surge motion.
The vertical location of its lower surface is denoted $Z(x,t)=-d+\real\{w(x)\ex^{-\ci \omega t}\}$.
The displacement function $w$ is decomposed as
\begin{equation}
w(x) 
= 
A_{H}+A_{P} x
+
\sum_{n=1}^{\infty}\xi_{n}w_{n}(x).
\end{equation}
Here $A_{H}$ and $A_{P}$  are complex-valued amplitudes of heave and pitch motions, respectively,
such that $a_{H}=\vert A_{H}\vert$ and $a_{P}=\vert A_{P}\vert$ .
(Here, pitch is defined with respect to rotations around the geometric centre of the disk's lower surface, 
as opposed to the centre of mass.
This has a negligible effect for the thin disk considered.)

The functions $w_{n}$ are the (symmetric) flexural modes of vibration 
and $\xi_{n}$ are the corresponding amplitudes.
\citet{ItaCra79} and \citet{MeySqu96} provide expressions for the flexural modes.
%\lu{Lucas: Mike's paper on circular floes in JGR circa 95 not 93}
For the incident wave and floe parameters studied here, 
the amplitudes $\xi_{n}\leq 5 \times 10^{-5}$\,mm, 
\ie the disk responds to the incident waves in its rigid modes only.

Disk motions are forced by differential pressures from the surrounding water due to hydrodynamics.
Air pressure is assumed to be constant, $p_{0}$ say.
Water pressure is modelled using the linearised version of Bernoulli's equation
\begin{equation}
p=p_{0}-\rho \frac{\partial \Phi}{\partial t} -\rho g z.
\end{equation}
Here $\Phi$ is the velocity potential of the water, \ie the water velocity field is the spatial gradient of $\Phi$.
Following potential-flow theory, it has been assumed the water is inviscid, homogeneous, incompressible and in irrotational motion.

Water motions are also assumed to be time harmonic, and the velocity potential is expressed 
\begin{equation} \label{eq:PF_Phi}
\Phi(x,y,z,t)=\real\left\{\frac{g}{\ci \omega}\phi(x,y,z)\ex^{-\ci \omega t}\right\}.
\end{equation}
The reduced (time independent) velocity potential, $\phi$, satisfies Laplace's equation throughout the water domain, \ie
\begin{equation}\label{eqn:laplace}
\nabla^{2}\phi+\frac{\partial^{2} \phi}{\partial z^{2}}=0
\quad
\textrm{where}
\quad
\nabla=(\partial/\partial x,\partial/\partial y),
\end{equation}
and a no-normal-flow condition on the floor of the basin, \ie
\begin{equation}
\frac{\partial\phi}{\partial z}=0
\quad
\textrm{on}
\quad
z=-h.
\end{equation}
The water is assumed to extend to infinity in all horizontal directions, \ie reflections from the basin boundaries are not considered.
In the far-field the velocity potential is composed of the incident-wave potential
\begin{equation} \label{eq:PF_phi_I}
\phi_{I}(x,y,z)
=
\frac{a\ex^{\ci k x}\cosh k(z+h)}{\cosh kh},
\end{equation}
and a geometrically decaying scattered-wave potential, which satisfies the Sommerfeld Radiation condition
\begin{equation}
\sqrt{r}
\left(
\partial_{r}
-\ci k
\right)
(\phi-\phi_{I})
\to 0
\quad
\textrm{as}
\quad
r\to\infty,
\end{equation}
where $r=\sqrt\{x^{2}+y^{2}\}$ is the radial coordinate.

The amplitudes of the waves and disk motions are assumed to be sufficiently small that linear theory is applicable.
Conditions on moving boundaries are therefore approximated by linearised conditions applied on the corresponding equilibrium boundaries.
The linearised free-surface condition is
\begin{equation}
\frac{\partial\phi}{\partial z}=\kappa\phi 
\quad
\textrm{on}
\quad
z=0,
\end{equation}
which holds at all horizontal points $(x,y)\notin\Omega$, where $\Omega=\{x,y: x^{2}+y^{2}=r^{2} \leq R^{2}\}$ 
is the projection of the floe in equilibrium onto the $xy$-plane.

Following Kirchhoff-Love thin-plate theory \citep{TimWoi59}, 
the linearised equation of motion in the $z$-direction is
\begin{equation}
(1-\kappa d)w+F\nabla^{4}w=\phi
\quad
\textrm{on} 
\quad
z=-d,
\end{equation}
and for $(x,y)\in\Omega$. Here 
%\begin{equation}
%F=\frac{Eh^{3}}{12\rho g(1-\nu^{2})}
%\end{equation}
$F=Eh^{3}/\{12\rho g(1-\nu^{2})\}$
is a scaled flexural rigidity of the disk, where $E=530$\,MPa is the Young's modulus, 
as measured by a three-point bending test on a strip of the Nycel.
The quantity $\nu=0.3$ is chosen as a typical value of Poisson's ratio for Nycel.
Note, the model is insensitive to the elastic parameters, $E$ and $\nu$, as the disk responds rigidly.
%\lu{Lucas: we need to look up some values for $E$ and $\nu$}
%\revmarcomm{Lucas: use $\backslash$, between a value and its units, \ie 530$\backslash$,MPa}

The motion of the disk and the velocity potential are also coupled via the linearised kinematic conditions
\begin{subequations}
\begin{equation}
\frac{\partial\phi}{\partial r}
=
\kappa \cos(\theta) X
\quad
\textrm{on}
\quad
(x,y)\in\delta\Omega,
%r=R,
%\;
%-\pi<\theta\leq \pi,
\;
-d<z<0,
\end{equation}
where $\theta=\tan(y/x)$ is the azimuthal coordinate, and
\begin{equation}
%\textrm{and}\quad
\frac{\partial\phi}{\partial z}
=
\kappa w
\quad
\textrm{on}
\quad
(x,y)\in\Omega,
\; z=-d.
\end{equation}
\end{subequations}
Lastly, the vertical displacements of the floe satisfy the linearised free-edge conditions
\begin{subequations}\label{eqns:freeedge}
\begin{equation}
\nabla^{2}w
-
(1-\nu)
\left(\frac{1}{r}\frac{\partial w}{\partial r}+\frac{1}{r^{2}}\frac{\partial^{2}w}{\partial \theta^{2}}\right)
=0
\quad
\textrm{and}
\end{equation}
\begin{equation}
\frac{\partial}{\partial r}
\nabla^{2}w
+
(1-\nu)
\frac{1}{r}
\frac{\partial}{\partial r}
\left(
\frac{1}{r}
\frac{\partial^{2}w}{\partial \theta^{2}}
\right)
=0
,
\end{equation}
\end{subequations}
both for $(x,y)\in\delta\Omega$. %$r=R$.

Equations (\ref{eqn:laplace})--(\ref{eqns:freeedge}) define a boundary value problem for the velocity potential, $\phi$.
The vertical displacement of the floe, $w$, is obtained as part of the solution.
At this juncture, the surge amplitude, $A_{S}$, is a parameter of the problem.

The velocity potential and vertical displacement function are decomposed as
\begin{equation}\label{eqn:phidecomp}
\phi = \hat{\phi} + A_{S}\check{\phi}
\quad
\textrm{and}
\quad
w = \hat{w} + A_{S}\check{w}.
\end{equation}
The functions $\hat{\phi}$ and $\hat{w}$ are the solutions of equations (\ref{eqn:laplace}) to (\ref{eqns:freeedge}) with the floe artificially restrained in surge, $A_{S}=0$.
The functions $\check{\phi}$ and $\check{w}$ are the solutions of equations (\ref{eqn:laplace}) to (\ref{eqns:freeedge}) with no incident wave forcing, $a=0$, but with the disk forced to oscillate with unit amplitude in surge, $A_{S}=1$.
These boundary value problems are solved using an eigenfunction matching method,
which is described  by \citet{LinMcI01} for general hydrodynamic problems.

The surge amplitude, $A_{S}$, for the full problem 
is found via the linearised equation of motion of the disk in the $x$-direction, 
which is
\begin{equation}\label{eqn:eomX}
-(\omega^2 m  + C_{m})  
A_{S}
=
f.
\end{equation}
The surge motion is forced by the pressure field around its edge created by the incident wave, \ie
\begin{equation} \label{eq:eomf}
f=
\rho g
\int_{-d}^{0}
\int_{-\pi}^{\pi}
R \cos(\theta)
[\hat{\phi}]_{(x,y)\in\delta\Omega}
\wrt \theta \wrt z.
\end{equation}
The complex quantity $C_{m}$ is defined by
\begin{equation} \label{eq:PF_Cm}
C_{m}=
\rho g
\int_{-d}^{0}
\int_{-\pi}^{\pi}
R \cos(\theta)
[\check{\phi}]_{(x,y)\in\delta\Omega}
\wrt\theta
\wrt z.
\end{equation}
The imaginary and real components of $C_{m}$ are known as the added mass and damping terms, respectively \citep{Mei83}.
The full solutions, $\phi$ and $w$, are obtained by substituting $A_{S}$ into equation  (\ref{eqn:phidecomp}).

% ========================================================================
\subsection{Long-wavelength regime}

\begin{figure}
  \centerline{\includegraphics[width=\textwidth]{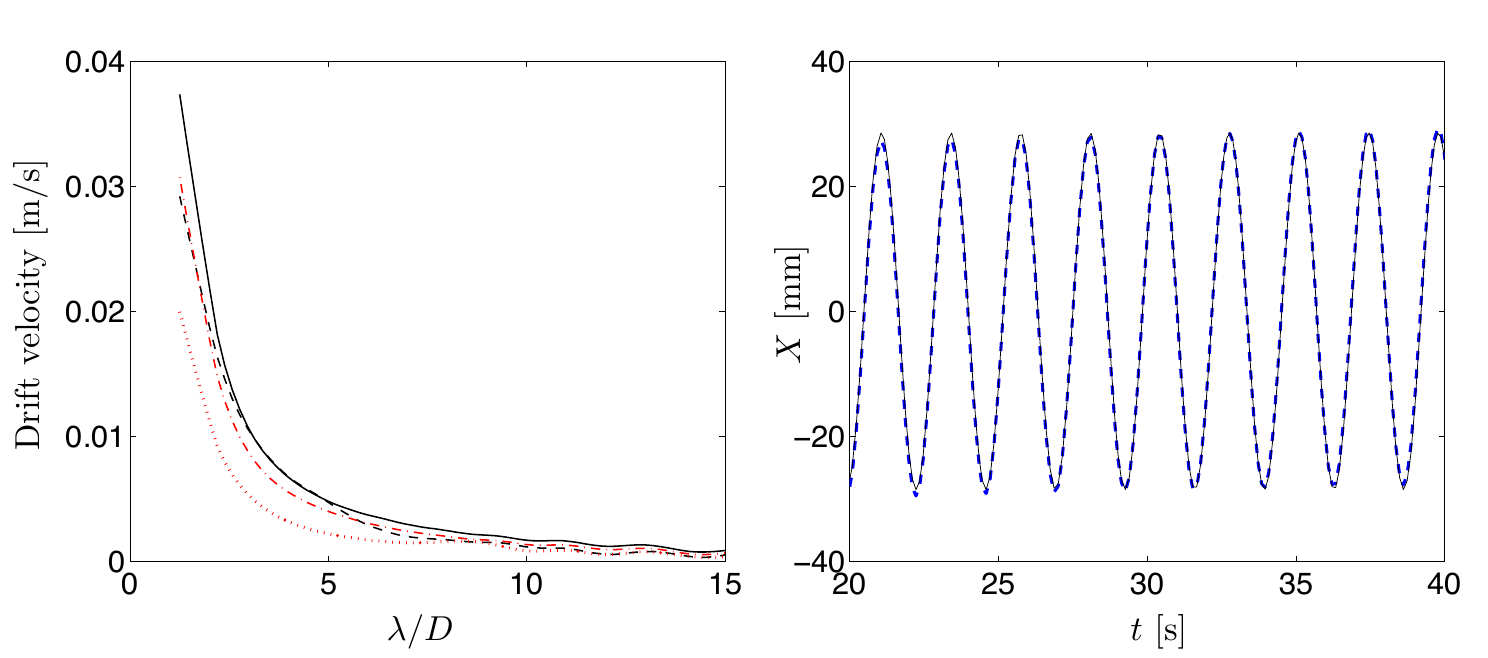}}
  \caption{Left-hand panel: drift velocity versus non-dimensional wavelength, as derived from the slope-sliding model. Here $c_m = c_d = 0$ (solid black curve), $c_m = 0$ and $c_d = 0.5$ (black dashed), $c_m = 0.1$ and $c_d = 0$ (red dot-dashed), $c_m = 0.1$ and $c_m = 0.5$ (dotted red curve). 
Right-hand panel: motion in the $x$-direction, calculated from the slope-sliding model, using the parameters $\lambda/D = 15$, and $c_m=c_d=0$. The surge component of the full numerical solution is represented by the blue dashed curve. Analytical long-wavelength solution is denoted by the solid black curve. }
  \label{fig:SS_drift}
\end{figure}

The slope-sliding model is valid in the long-wavelength regime.
The left-hand panel of figure~\ref{fig:SS_drift} shows the drift velocity predicted by the slope-sliding model tends to zero as the incident wavelength becomes large.
The figure shows this holds for zero and non-zero drag and added mass.
The slope-sliding model, therefore, predicts the horizontal motion of the disk is dominated by oscillatory surge in the large-wavelength regime.

For zero drag, which gives the best fit to the experimental data \citep{MeyEA14},
the slope-sliding model is
\begin{equation} \label{eq:SS_lim}
%\begin{array}{rcl}
\displaystyle
(1+c_{m})
\frac{dV}{dt} 
 = 
 %\displaystyle
 - g \left[\frac{\partial \eta}{\partial x} \right]_{x=X}
  %\\[12pt]
 =
 -gak \cos(kX-\omega t)
 \\[8pt]
\end{equation}
%Note that the wave profile, $\eta = a \sin{(k x - \omega t)}$.
The small drift is translated into the condition $\vert k X\vert \ll 1$.
The slope-sliding model is, therefore, manipulated into the form
\begin{equation} \label{eq:SS_lim2}
%\begin{aligned}
(1+c_m) 
\frac{dV}{dt} 
 = 
 - g a k \cos{\omega t} + \order (k X),
 %\left( \left[ 1 + O(k^2 x^2) \right] \cos{\omega t} + O(k x) \sin{\omega t} \right)  .
%\end{aligned}
\end{equation}
which is integrated to give
%Integrating (\ref{eq:SS_lim2}) with respect to time gives 
\begin{equation} \label{eq:SS_lim3}
%\begin{aligned}
 V(t) 
 = \frac{-g a k\sin(\omega t)}{\omega(1+c_{m})}  + \order (k X)  .
%\end{aligned}
\end{equation}
%The displacement of the floe, $\mathcal{X}$ is therefore
Thus, to leading order in $kX$, the displacement of the disk is
\begin{equation} \label{eq:SS_lim4}
%\begin{aligned}
 X(t) 
 = \frac{g a k\cos(\omega t) }{\omega^2 (1+c_m)} 
 = \frac{a \coth(k h) \cos(\omega t)}{1+c_{m}}  ,
%\end{aligned}
\end{equation}
where the final expression is derived using the dispersion relation.
The slope-sliding model, therefore, predicts the surge amplitude $a_{S}=a\coth(kh)/(1+c_{m})$ in the long-wavelength regime.
The right-hand panel of figure~\ref{fig:SS_drift} shows the leading-order displacement (\ref{eq:SS_lim4})  accurately approximates the surge motion of the full solution for $\lambda/D = 15$ and $c_m=c_d=0$.

The long-wavelength solution for the potential-flow model can also be solved analytically by considering the horizontal equation of motion in equations (\ref{eqn:eomX}) to (\ref{eq:PF_Cm}). 
In long waves, the velocity potential is approximately equal to the incident potential $\phi_I$, as with equation (\ref{eq:PF_phi_I}), due to the wave field remaining unaltered by the floe.

In (\ref{eq:PF_phi_I}), the term $\ex^{\ci k x} = \ex^{ \ci k R \cos{\theta} }$ can be expressed as $ 1 + \ci k R \cos{\theta} + \order (k R)^2$ through Taylor expansion.
Since $k R \ll 1$ in long waves, $\order (k R)^2$ may be disregarded.
Substituting the remaining terms into equation (\ref{eq:eomf}) gives
\begin{equation} 
\label{eq:PF_lim2}
%\begin{array}{rcl}
%\displaystyle
f = \rho g  \int_{-d}^{0} \int_{-\pi}^{\pi} 
 a R \cos{\theta} (1 + \ci k R \cos{\theta} ) \frac{\cosh{k(z+h)}}{\cosh{k h}}
\wrt \theta \wrt z  
%\\[0.3cm]
%\displaystyle
\approx \ci a g k ~ \pi R^2 d \rho
%\end{array}
\end{equation}
as $k$ tends to zero (\ie in long waves). Note that $\pi R^2 d \rho = m$ as required by Archimedes' principle.
Equating this with the left-hand side of (\ref{eqn:eomX}) gives 
\begin{equation} \label{eq:PF_A_S}
A_S = \frac{- \ci a g k m}  {\omega^2 m + C_m } ,
\end{equation}
hence the surge amplitude is
\begin{equation}
a_S = |A_S| =  \frac{ a \coth{kh} }{| 1 + \hat{C}_m |}
\end{equation}
after applying the dispersion relation.
In the above, $\hat{C}_m = C_p + \ci C_a$ where $C_p$ is the damping coefficient and $C_a$ is the added mass coefficient nondimensionalised with respect to $\omega^2 m$.
These two coefficients can be calculated using equation (\ref{eq:PF_Cm}).

Figure \ref{fig:PF_Cm} shows the value of the real and imaginary parts of $\hat{C}_m$ for increasing wavelengths, calculated using dimensions of the disks described in \S~\ref{sec:Exp_method}.
In the long-wavelength limit, both the added mass and damping coefficients tend to zero.
At the largest wavelengths considered, \ie $\lambda / D = 12.3$, $C_a = C_p = \order(10^{-3})$.
Based on this, the assumption is that $\hat{C}_m \approx 0$ in long waves, and therefore the surge amplitude will tend to $a_S = a \coth{kh}$.

\begin{figure}
  \centerline{\includegraphics[width=\textwidth]{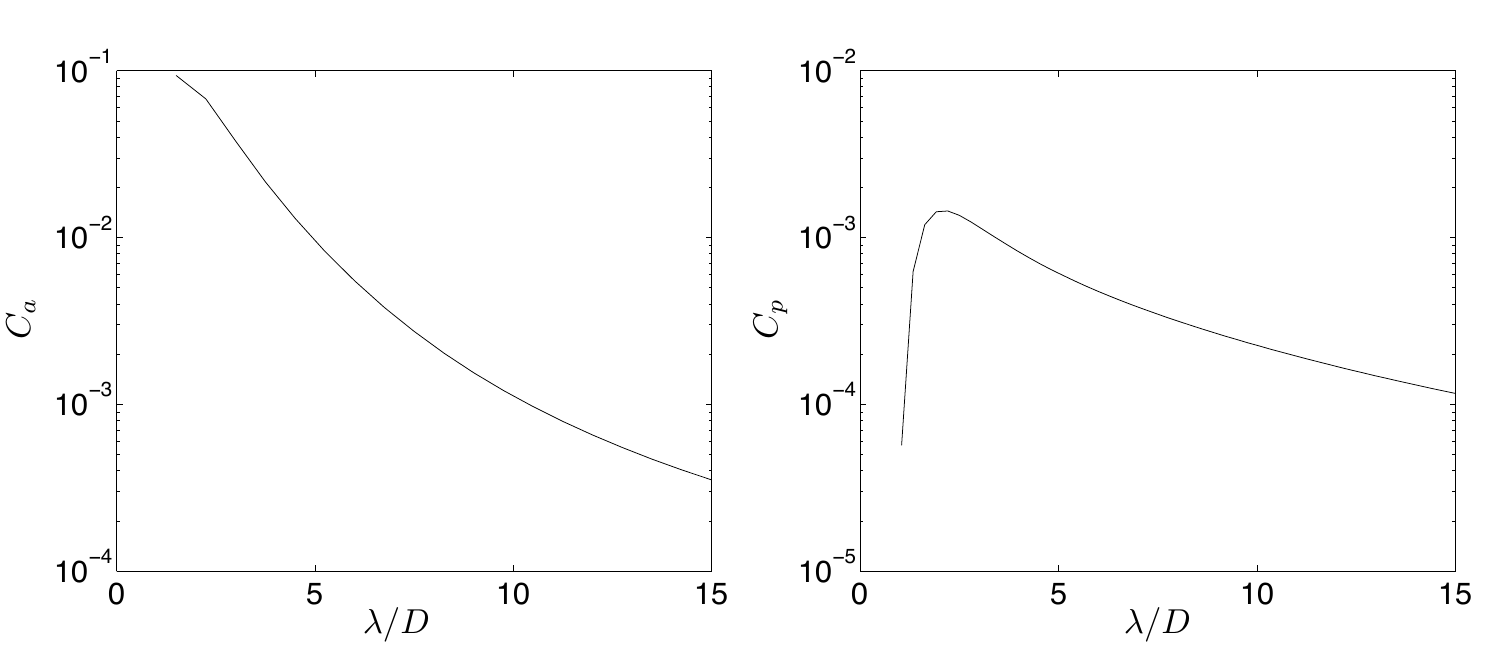}}
  \caption{Added mass and damping coefficients (left and right-hand panels, respectively), calculated from the potential-flow model, as a function of non-dimensional wavelength.}
  \label{fig:PF_Cm}
\end{figure}

%%% RESULTS %%%%

\section{RAOs: potential-flow/thin-plate model and data comparison} \label{sec:Res}

\begin{figure}
  \centerline{\includegraphics[width=\textwidth]{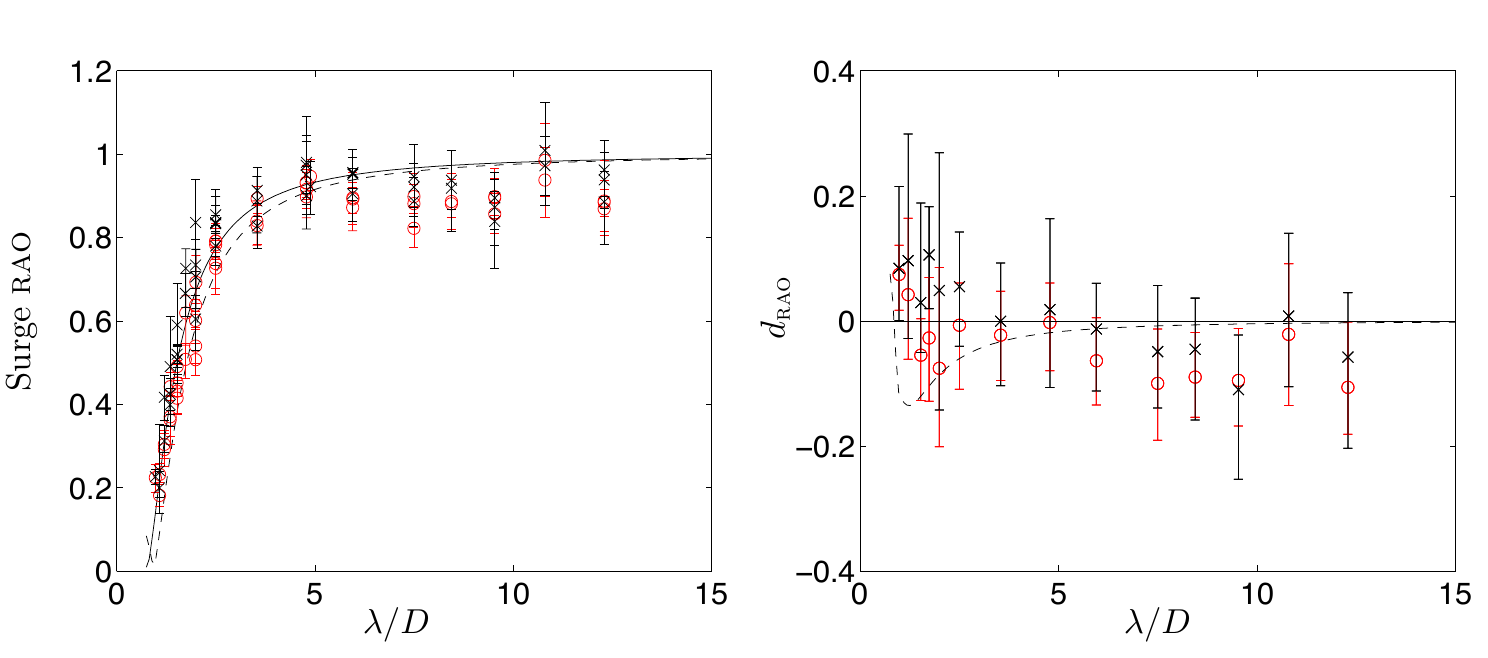}}
  \caption{Left-hand panel: comparison between potential-flow models (disk: solid curve, 2-D: dashed) and experimental data for surge as per figure \ref{fig:Exp_Res_Ice1Ice2}. Error bars show the range of error for each data point. Right-hand panel: experimental data for each disk are grouped according to wavelengths, and the means for each group are plotted in terms of its difference ($d_{\textsc{rao}}$) from the disk disk model. Error bars represent the overall maxima and minima of each group. The comparison with the 2-D potential-flow model (dashed) is also shown.}
\label{fig:Res_PFvExp_surge}
\end{figure}

\begin{figure}
  \centerline{\includegraphics[width=\textwidth]{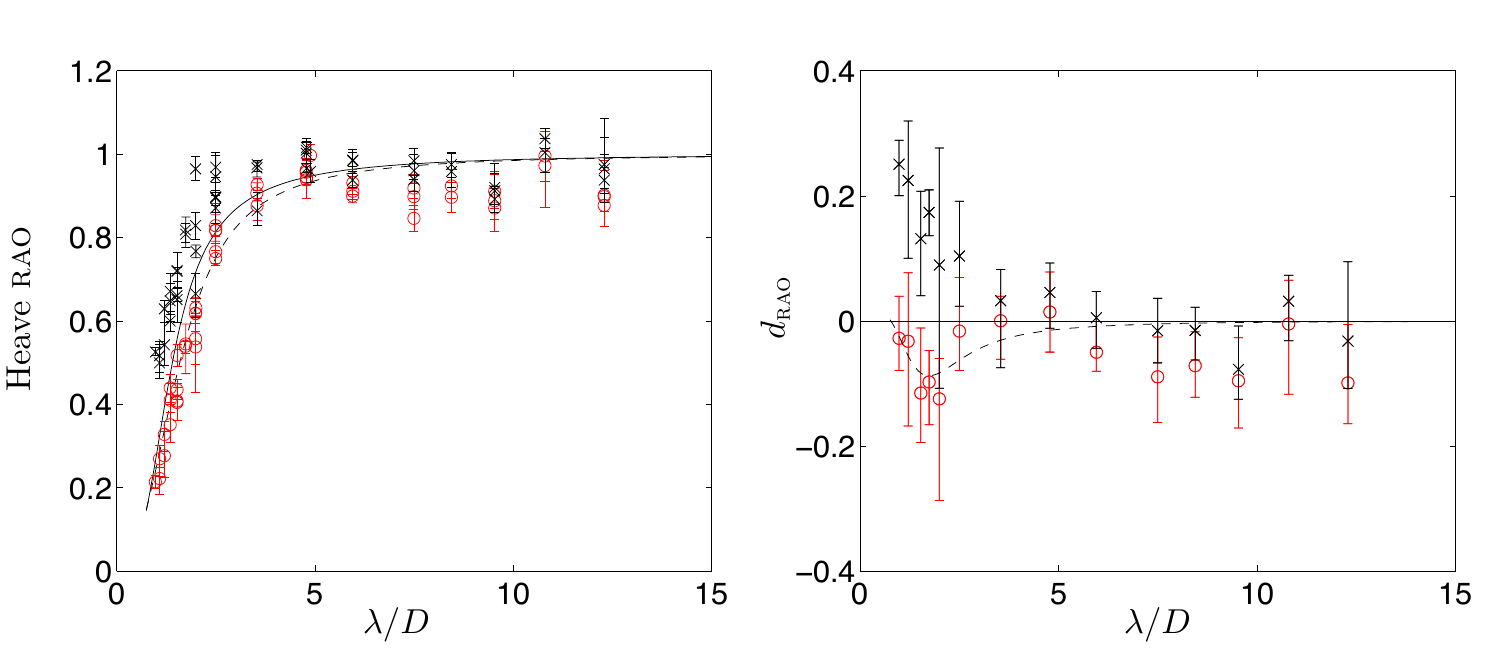}}
  \caption{As per figure \ref{fig:Res_PFvExp_surge} but for heave.}
\label{fig:Res_PFvExp_heave}
\end{figure}

\begin{figure}
  \centerline{\includegraphics[width=\textwidth]{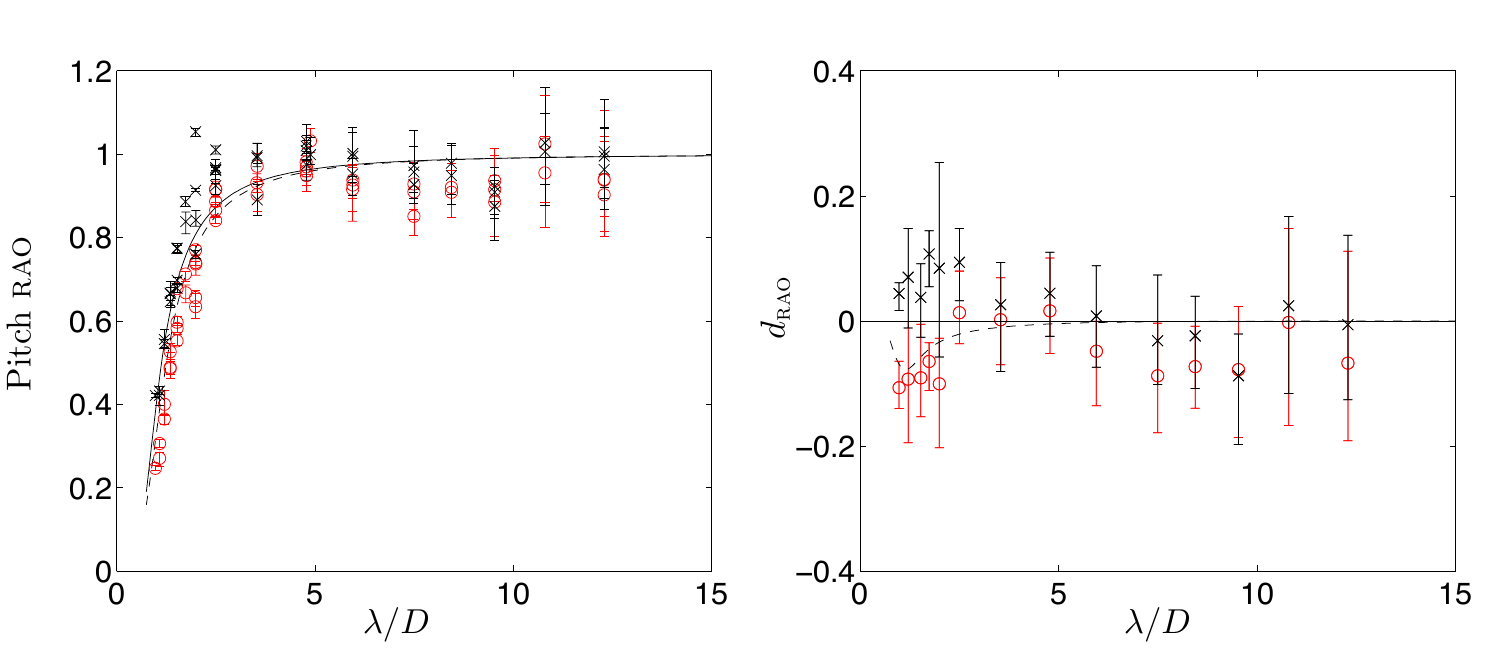}}
  \caption{As per figure \ref{fig:Res_PFvExp_surge} but for pitch.}
\label{fig:Res_PFvExp_pitch}
\end{figure}

{
The left-hand panels of figures~\ref{fig:Res_PFvExp_surge} to \ref{fig:Res_PFvExp_pitch}
show the \textsc{rao}s predicted of the potential-flow/thin-plate model for surge, heave and pitch, respectively,
as functions of non dimensional incident wavelength.
Predictions are shown for the disk model outlined in \S~\ref{sec:PF} and a simplified two-dimensional model.
%\citep{Ben&Chu11}.
The model predictions are overlaid on the \textsc{rao}s calculated from the experimental data,
as given in figure~\ref{fig:Exp_Res_Ice1Ice2}.
Error bars are included for the experimental results here. 
As in figure~\ref{fig:Exp_RAO_Lin}, 
the upper and lower limits are calculated using maximum and minimum amplitudes for each test. 
The right-hand panels show the corresponding differences between the data and the disk model.

Visually, the potential-flow/thin-plate models predict the \textsc{rao}s measured during the experiments with pleasing accuracy.
The right-hand panels of figures~\ref{fig:Res_PFvExp_surge} to \ref{fig:Res_PFvExp_pitch} show the differences between the disk model and the experimental data.
Here, experimental data are divided into 14 groups, with each group corresponding to a particular wavelength.
The means for each group are plotted (separately for Disks B and NB) in terms of its difference from the disk model.
Error bars on the right-hand panels represent the overall maxima and minima for each group.
Differences between the means and the disk model are generally very small (less than 30\%) across the range of wavelengths considered.
For Disk B, the average difference is greatest for heave at 6.8\%, and least for surge at 1.2\%.
For Disk NB, the average difference is greatest for heave and pitch at -5.6\%, and least for surge at -3.9\%.
The general trend

Differences between the two-dimensional and disk models are evident only for $\lambda/D<10$, approximately.
The differences are generally very small across the range of wavelengths considered.
{The differences are greatest for surge at 4\%, and least for pitch at 1.1\%.}
However, the two-dimensional model predicts a phase change in surge at {$\lambda/D\approx 6$}, 
which is not predicted by the disk model.
This implies the models will be dissimilar for shorter wavelengths.
%...difference at $\lambda/D < 3$ and $> 3$ etc.

%\lu{Lucas: comment of the right-hand panels}

Figure~\ref{fig:Exp_PF_RAO_Lin} shows the deviation between the disk model and the experimental data from the second test matrix.
The deviation is presented in a similar fashion to figure~\ref{fig:Exp_RAO_Lin}, but with the model predictions replacing the mean,
The model generally underpredicts the measured surge \textsc{rao}s for  the longer wavelength, $\lambda/D=2.50$.
There are two exceptions, both for Disk~NB.
%The deviation has mean \lu{??? (lucas to calculate)} for Disk~B and \lu{???} for Disk~B.
%The model predicts the surge of the disk without a barrier more accurately than the disk with a barrier.
%\lu{Lucas: did overwash occur in for these tests?}
The model also underpredicts the surge motion of Disk~B for the shorter wavelength, $\lambda/D=1.78$.
%The deviateion for Disk~NB has mean \lu{???}.
The model underpredicts the surge of Disk~B for the lowest steepness only.
The decrease  in surge \textsc{rao}s for Disk~NB with increasing steepness results in the model overpredicting 
the surge motion. %, with the deviation from the \textsc{rao} up to \lu{???} for the steepest wave considered. 

\begin{figure}
  \centerline{\includegraphics[width=\textwidth]{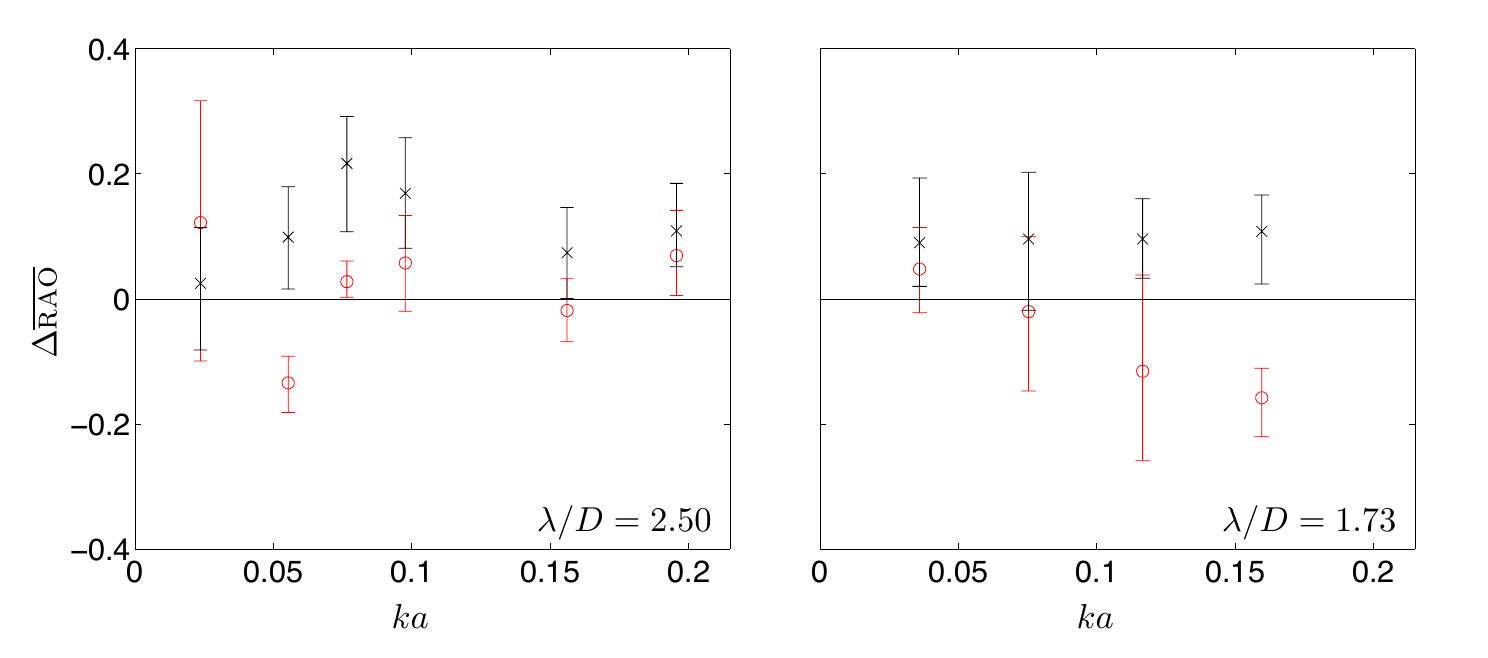}}
\caption{As per figure \ref{fig:Exp_RAO_Lin}, but here, deviation is calculated with respect to the potential-flow model (horizontal solid line).}
\label{fig:Exp_PF_RAO_Lin}
\end{figure}

% ############################################################################
\section{Summary and conclusions}

%\lu{Needs work --- leave for now.}

{
Wave basin experiments were conducted to determine the hydrodynamic responses of an ice floe in regular waves, over a range of wave conditions.
Wave amplitudes between 2.5 and 50\,mm, and wave frequencies between 0.5 and 2.0\, Hz were tested.
Scale models of typical small-medium sized floes were used.
The floes, which were constructed from Nycel, measured 400\,mm in diameter and 15\,mm in thickness.
An edge barrier was used to determine the effect of overwashing.
Results for the single floe experiments showed that: 
(a) surge, heave and pitch motions are, in general, linear with respect to the incident wave amplitude; 
(b) there was a reduction in all three motions in the short-wavelength regime as a result of scattering; 
(c) overwashing occurs in the short-wavelength regime and is dependent on the incident wave amplitude; 
(d) overwashing appears to decrease the response of all three motions 
--- the greatest reduction occurred in heave, while surge was least affected.

Two theoretical models were presented --- a slope-sliding model and a linear potential-flow/thin-plate model.
Comparisons with experimental data showed that the potential-flow/thin-plate model provided good agreement across the full range of wavelengths tested.
The slope-sliding model was only accurate in the long wavelength regime.
%In the potential-flow model, elasticity was considered, however it was shown to be insignificant within the bounds of the test conditions.

A simplified 2-D potential-flow model was also proposed, by disregarding the lateral dimension.
Although there were minimal differences in the responses predicted by the simplified and full models, the latter had a slightly better agreement with the experimental data.
The two-dimensional model, however, was still considered to be reasonably accurate.
}

\section*{Acknowledgements}

The Australian Research Council funds LJY's PhD Scholarship and provides funding support to LGB (DE130101571).
The Australian Antarctic Science Program also provides support for LGB (Project 4123). 
The U.S.\ Office of Naval Research provides funding support to MHM.
The Australian Maritime College funded the experiments.
%The authors thank XX and YY for technical support during the experiments.

%\bibliographystyle{plainnat} 
%\bibliography{../../References/References,../Bibli_luke}

     % References
%\bibliography{../02/references,/Users/a1612881/Dropbox/Documents/Bibli/Bibli,/Users/a1612881/Dropbox/Documents/Bibli/MyPublications}

\begin{thebibliography}{13}
\expandafter\ifx\csname natexlab\endcsname\relax\def\natexlab#1{#1}\fi

  \bibitem[Bennetts {\em et~al.\/}(2010)Bennetts, Peter, Squire \&
  Meylan]{Benetal10}
{\sc Bennetts, L.~G., Peter, M.~A., Squire, V.~A. \& Meylan, M.~H.} 2010 A
  three-dimensional model of wave attenuation in the marginal ice zone. {\em J.
  Geophys. Res.\/} {\bf 115}, C12043.

\bibitem[Bennetts \& Squire(2012)]{Ben&Squ12a}
{\sc Bennetts, L.~G. \& Squire, V.~A.} 2012 On the calculation of an
  attenuation coefficient for transects of ice-covered ocean. {\em Proc. R.
  Soc. Lond. A\/} {\bf 468}~(2137), 136--162.

\bibitem[Bennetts \& Williams(2015)]{BenWil15}
{\sc Bennetts, L.~G. \& Williams, T.~D.} 2015 Water wave transmission by an
  array of floating disks. {\em Proc. R. Soc. A\/} {\bf 470}~(2173).
  

\bibitem[Dai {\em et~al.\/}(2004)Dai, Shen, Hopkins \& Ackley]{Daietal04}
{\sc Dai, M., Shen, H.~H., Hopkins, M.~A. \& Ackley, S.~F.} 2004 Wave rafting
  and the equilibrium pancake ice cover thickness. {\em J. Geophys. Res.\/}
  {\bf 109}~(C7), C07023.


\bibitem[Grotmaack \& Meylan(2006)]{GroMey06}
{\sc Grotmaack, R. \& Meylan, M.~H.} 2006 Wave forcing on small floating
  bodies. {\em J. Waterway, Port, Coastal, Ocean Eng.\/} {\bf 132}~(3),
  192--198.

\bibitem[Harms(1987)]{Har87}
{\sc Harms, V.~W.} 1987 Steady wave-drift of modeled ice floes. {\em J.
  Waterway, Port, Coastal, Ocean Eng.\/} {\bf 113}, 606--622.
  
\bibitem[Huang {\em et~al.\/}(2011)Huang, Law \& Huang]{Huaetal11}
{\sc Huang, G., Law, A. W.-K. \& Huang, Z.} 2011 Wave-induced drift of small
  floating objects in regular waves. {\em Ocean Engineering\/} {\bf 38}~(4),
  712 -- 718.

\bibitem[Isaacson(1982)]{Isa82}
{\sc Isaacson, M.} 1982 Fixed and floating axisymmetric structures in waves.
  {\em J. Waterw. Port. Coastal and Ocean Div.\/} pp. 180--199.

\bibitem[Itao \& Crandall(1979)]{ItaCra79}
{\sc Itao, K. \& Crandall, S.~H.} 1979 Natural modes and natural frequencies of
  uniform, circular, free-edge plates. {\em J. Appl. Mech.\/} {\bf 46}~(2),
  448--453.
  
 \bibitem[Kohout \& Meylan(2008)]{Koh&Mey08a}
{\sc Kohout, A.~L. \& Meylan, M.~H.} 2008 An elastic plate model for wave
  attenuation and ice floe breaking in the marginal ice zone. {\em J. Geophys.
  Res.\/} {\bf 113}~(C09016).

\bibitem[Langhorne {\em et~al.\/}(2001)Langhorne, Squire \& Haskell]{Lanetal01}
{\sc Langhorne, P.~J., Squire, V.~A. \& Haskell, T.~G.} 2001 Lifetime
  estimation for a fast ice sheet subjected to ocean swell. {\em Annals
  Glaciol.\/} {\bf 33}, 333--338.
  
\bibitem[Linton \& McIver(2001)]{LinMcI01}
{\sc Linton, C.~M. \& McIver, P.} 2001 {\em Handbook of Mathematical Techniques
  for Wave/Structure Interactions\/}. Chapman \& Hall/CRC.

\bibitem[Marchenko(1999)]{Mar99}
{\sc Marchenko, A.~V.} 1999 The floating behaviour of a small body acted upon
  by a surface wave. {\em J. Appl. Math. Mech.\/} {\bf 63}~(3), 471--478.

\bibitem[Martin \& Becker(1987)]{Mar&Bec87}
{\sc Martin, S. \& Becker, P.} 1987 High-frequency ice floe collisions in the
  {G}reenland {S}ea during the 1984 marginal ice zone experiment. {\em J.
  Geophys. Res.\/} {\bf 92}~(C7), 7071--7084.

\bibitem[Massom \& Stammerjohn(2010)]{Mas&Sta10}
{\sc Massom, R. \& Stammerjohn, S.} 2010 Antarctic sea ice variability:
  Physical and ecological implications. {\em Polar Sci.\/} {\bf 4}, 149--458.

\bibitem[Masson \& LeBlond(1989)]{Mas&LeB89}
{\sc Masson, D. \& LeBlond, P.} 1989 Spectral evolution of wind-generated
  surface gravity waves in a dispersed ice field. {\em J. Fluid Mech.\/} {\bf
  202}, 111--136.

\bibitem[{McGovern} \& Bai(2014)]{McG&Bai14}
{\sc {McGovern}, D. \& Bai, W.} 2014 Experimental study on kinematics of sea
  ice floes in regular waves. {\em Cold Reg. Sci. Technol.\/} {\bf 103},
  15--30.

\bibitem[{McK}enna \& Croker(1990)]{McK&Cro90}
{\sc {McK}enna, R.~F. \& Croker, G.~B.} 1990 Wave energy and floe collisions in
  marginal ice zones. In {\em Ice Technology for Polar Regions, proceedings of
  the Second International Conference on Ice Technology\/} (ed. T.~K.~S.
  Murphy), pp. 33--41. Southampton: Computational Mechanics Publications.
  
  \bibitem[Mei(1983)]{Mei83}
{\sc Mei, C.~C.} 1983 {\em The Applied Dynamics of Ocean Surface Waves\/}. John
  Wiley \& Sons, Inc.

\bibitem[Meylan(2002)]{Mey02}
{\sc Meylan, M.~H.} 2002 Wave response of an ice floe of arbitrary geometry.
  {\em J. Geophys. Res.\/} {\bf 107}~(C1), 5--1--5--11.

\bibitem[Meylan \& Squire(1996)]{MeySqu96}
{\sc Meylan, M.~H. \& Squire, V.~A.} 1996 Response of a circular ice floe to
  ocean waves. {\em J. Geophys. Res.\/} {\bf 101}~(C4), 8869--8884.

\bibitem[Meylan {\em et~al.\/}(2015)Meylan, Yiew, Bennetts, French \&
  Thomas]{MeyEA14}
{\sc Meylan, M.~H., Yiew, L.~J., Bennetts, L.~G., French, B.~J. \& Thomas,
  G.~A.} 2015 Surge motion of an ice floe in waves: comparison of theoretical
  and experimental models. {\em Ann. Glaciol.\/} {\bf 56}~(69), 107--111.

\bibitem[Meylan \& Squire(1996)]{Mey&Squ96}
{\sc Meylan, M.~H. \& Squire, V.~A.} 1996 Response of a circular ice floe to
  ocean waves. {\em J. Geophys. Res.\/} {\bf 101}, 8869--8884.

\bibitem[Meylan {\em et~al.\/}(1997)Meylan, Squire \& Fox]{Meyetal97}
{\sc Meylan, M.~H., Squire, V.~A. \& Fox, C.} 1997 Towards realism in modeling
  ocean wave behavior in marginal ice zones. {\em J. Geophys. Res.\/} {\bf
  102}~(C10), 22981--22991.

\bibitem[Meylan {\em et~al.\/}(2015)Meylan, Yiew, Bennetts, French \&
  Thomas]{Meyetal15a}
{\sc Meylan, M.~H., Yiew, L.~J., Bennetts, L.~G., French, B.~J. \& Thomas,
  G.~T.} 2015 Surge motion of an ice floe in waves: comparison of theoretical
  and experimental models. {\em Annals Glaciol.\/} {\bf 56}~(69), 107---111.

\bibitem[Montiel {\em et~al.\/}(2013{\natexlab{{\em a\/}}})Montiel, Bennetts,
  Squire, Bonnefoy \& Ferrant]{Monetal13b}
{\sc Montiel, F., Bennetts, L.~G., Squire, V.~A., Bonnefoy, F. \& Ferrant, P.}
  2013{\natexlab{{\em a\/}}} Hydroelastic response of floating elastic disks to
  regular waves. {P}art 2: {M}odal analysis. {\em J. Fluid Mech.\/} {\bf 723},
  629--652.

\bibitem[Montiel {\em et~al.\/}(2013{\natexlab{{\em b\/}}})Montiel, Bonnefoy,
  Ferrant, Bennetts, Squire \& Marsault]{Monetal13a}
{\sc Montiel, F., Bonnefoy, F., Ferrant, P., Bennetts, L.~G., Squire, V.~A. \&
  Marsault, P.} 2013{\natexlab{{\em b\/}}} Hydroelastic response of floating
  elastic disks to regular waves. {P}art 1: {W}ave tank experiments. {\em J.
  Fluid Mech.\/} {\bf 723}, 604--628.

\bibitem[Prinsenberg \& Peterson(2011)]{Pri&Pet11}
{\sc Prinsenberg, S.~J. \& Peterson, I.~K.} 2011 Observing regional-scale
  pack-ice decay processes with helicopter-borne sensors and moored
  upward-looking sonars. {\em Annals Glaciol.\/} {\bf 52}, 35--42.

\bibitem[Rumer {\em et~al.\/}(1979)Rumer, Chrissmas \& Wake]{RumEA79}
{\sc Rumer, R.~R., Chrissmas, R.~D. \& Wake, A.} 1979 {\em Ice Transport in
  Great Lakes\/}. {\em Water Resources and Environmental Engineering
  Research\/} 79-3. NY State Univ. at Buffalo.

\bibitem[Shen \& Zhong(2001)]{SheZho01}
{\sc Shen, H.~H. \& Zhong, Y.} 2001 Theoretical study of drift of small rigid
  floating objects in wave fields. {\em J. Waterway, Port, Coastal, Ocean
  Eng.\/} {\bf 127}, 343--351.
  
  \bibitem[Shen \& Ackley(1991)]{She&Ack91}
{\sc Shen, H.~H. \& Ackley, S.~F.} 1991 A one-dimensional model for
  wave-induced ice-floe collisions. {\em Annals Glaciol.\/} {\bf 15}, 87--95.

\bibitem[Squire \& Moore(1980)]{Squ&Mor80}
{\sc Squire, V.~A. \& Moore, S.~C.} 1980 Direct measurement of the attenuation
  of ocean waves by pack ice. {\em Nature\/} {\bf 283}, 365--368.

\bibitem[Squire(2007)]{Squ07}
{\sc Squire, V.~A.} 2007 Of ocean waves and sea-ice revisited. {\em Cold Reg.
  Sci. Technol.\/} {\bf 49}, 110--133.

\bibitem[Squire {\em et~al.\/}(1995)Squire, Dugan, Wadhams, Rottier \&
  Liu]{Squetal95}
{\sc Squire, V.~A., Dugan, J.~P., Wadhams, P., Rottier, P.~J. \& Liu, A.~K.}
  1995 Of ocean waves and sea ice. {\em Annu. Rev. Fluid Mech.\/} {\bf 27},
  115--168.

\bibitem[Squire(2007)]{Squ07}
{\sc Squire, V.~A.} 2007 Of ocean waves and sea-ice revisited. {\em Cold
  Regions Science and Technology\/} {\bf 49}~(2), 110 -- 133.

\bibitem[Timoshenko \& Woinowsky-Krieger(1959)]{TimWoi59}
{\sc Timoshenko, S. \& Woinowsky-Krieger, S.} 1959 {\em Theory of Plates and
  Shells\/}, 2nd edn. New York: McGraw-Hill.

\bibitem[Wadhams(1983)]{Wad83}
{\sc Wadhams, P.} 1983 A mechanism for the formation of ice edge bands. {\em J.
  Geophys. Res.\/} {\bf 88}~(C5), 2813--2818.

\bibitem[Wadhams {\em et~al.\/}(1979)Wadhams, Gill \& Linden]{Wadetal79a}
{\sc Wadhams, P., Gill, A.~E. \& Linden, P.~F.} 1979 Transects by submarine of
  the {E}ast {G}reenland {P}olar {F}ront. {\em Deep Sea Res.\/} {\bf 26}~(A),
  1311--1327.

\bibitem[Williams {\em et~al.\/}(2013{\natexlab{{\em a\/}}})Williams, Bennetts,
  Dumont, Squire \& Bertino]{Wiletal13a}
{\sc Williams, T.~D., Bennetts, L.~G., Dumont, D., Squire, V.~A. \& Bertino,
  L.} 2013{\natexlab{{\em a\/}}} Wave-ice interactions in the marginal ice
  zone. {P}art 1: {T}heoretical foundations. {\em Ocean Model.\/} {\bf 71},
  81--91.

\bibitem[Williams {\em et~al.\/}(2013{\natexlab{{\em b\/}}})Williams, Bennetts,
  Dumont, Squire \& Bertino]{Wiletal13b}
{\sc Williams, T.~D., Bennetts, L.~G., Dumont, D., Squire, V.~A. \& Bertino,
  L.} 2013{\natexlab{{\em b\/}}} Wave-ice interactions in the marginal ice
  zone. {P}art 2: {N}umerical implementation and sensitivity studies along 1d
  transects of the ocean surface. {\em Ocean Model.\/} {\bf 71}, 92--101.
  
\end{thebibliography}

\end{document}